\documentclass[%
reprint,
superscriptaddress,
nofootinbib,
 amsmath,amssymb,
aps,
prx,
floatfix,
]{revtex4-2}

\usepackage{graphicx}
\usepackage{dcolumn}
\usepackage{bm}
\usepackage[hypertexnames=false]{hyperref}
\usepackage[all]{hypcap}    
\usepackage{siunitx}
\usepackage{braket}
\usepackage{rotating}
\usepackage{algorithm} 
\usepackage{algpseudocode} 
\usepackage{xfrac}
\usepackage{subcaption}

\begin{document}
\title{Telecom-band quantum interference of frequency-converted photons from remote detuned NV centers}

\author{A. J. Stolk}\altaffiliation{These authors contributed equally to this work}
\affiliation{\vspace{0.5em}QuTech, Delft University of Technology, 2628 CJ Delft, The Netherlands}
\affiliation{\vspace{0.5em}Kavli Institute of Nanoscience, Delft University of Technology, 2628 CJ Delft, The Netherlands}
\author{K. L. van der Enden}\altaffiliation{These authors contributed equally to this work}
\affiliation{\vspace{0.5em}QuTech, Delft University of Technology, 2628 CJ Delft, The Netherlands}
\affiliation{\vspace{0.5em}Kavli Institute of Nanoscience, Delft University of Technology, 2628 CJ Delft, The Netherlands}
\author{M.-C. Roehsner}
\affiliation{\vspace{0.5em}QuTech, Delft University of Technology, 2628 CJ Delft, The Netherlands}
\affiliation{\vspace{0.5em}Kavli Institute of Nanoscience, Delft University of Technology, 2628 CJ Delft, The Netherlands}

\author{A. Teepe}
\affiliation{\vspace{0.5em}QuTech, Delft University of Technology, 2628 CJ Delft, The Netherlands}
\affiliation{\vspace{0.5em}Kavli Institute of Nanoscience, Delft University of Technology, 2628 CJ Delft, The Netherlands}

\author{S.O.F. Faes}
\affiliation{\vspace{0.5em}QuTech, Delft University of Technology, 2628 CJ Delft, The Netherlands}
\affiliation{\vspace{0.5em}Kavli Institute of Nanoscience, Delft University of Technology, 2628 CJ Delft, The Netherlands}

\author{S. Cadot}
\affiliation{\vspace{0.5em}QuTech, Delft University of Technology, 2628 CJ Delft, The Netherlands}

\author{J. van Rantwijk}
\affiliation{\vspace{0.5em}QuTech, Delft University of Technology, 2628 CJ Delft, The Netherlands}

\author{I. te Raa}
\affiliation{\vspace{0.5em}QuTech, Delft University of Technology, 2628 CJ Delft, The Netherlands}

\author{R.A.J. Hagen}
\affiliation{\vspace{0.5em}QuTech, Delft University of Technology, 2628 CJ Delft, The Netherlands}
\affiliation{\vspace{0.5em}Netherlands Organisation for Applied Scientific Research (TNO), P.O. Box 155, 2600 AD Delft, The Netherlands}
\author{A.L. Verlaan}
\affiliation{\vspace{0.5em}QuTech, Delft University of Technology, 2628 CJ Delft, The Netherlands}
\affiliation{\vspace{0.5em}Netherlands Organisation for Applied Scientific Research (TNO), P.O. Box 155, 2600 AD Delft, The Netherlands}
\author{B. Biemond}
\affiliation{\vspace{0.5em}QuTech, Delft University of Technology, 2628 CJ Delft, The Netherlands}
\affiliation{\vspace{0.5em}Netherlands Organisation for Applied Scientific Research (TNO), P.O. Box 155, 2600 AD Delft, The Netherlands}
\author{A. Khorev}
\affiliation{\vspace{0.5em}QuTech, Delft University of Technology, 2628 CJ Delft, The Netherlands}
\affiliation{\vspace{0.5em}Netherlands Organisation for Applied Scientific Research (TNO), P.O. Box 155, 2600 AD Delft, The Netherlands}

\author{R. Vollmer}
\affiliation{\vspace{0.5em}QuTech, Delft University of Technology, 2628 CJ Delft, The Netherlands}
\affiliation{\vspace{0.5em}Netherlands Organisation for Applied Scientific Research (TNO), P.O. Box 155, 2600 AD Delft, The Netherlands}
\author{M. Markham}
\affiliation{\vspace{0.5em}Element Six Innovation, Fermi Avenue, Harwell Oxford, Didcot, Oxfordshire OX11 0QR, UK}
\author{A. M. Edmonds}
\affiliation{\vspace{0.5em}Element Six Innovation, Fermi Avenue, Harwell Oxford, Didcot, Oxfordshire OX11 0QR, UK}
\author{J.P.J. Morits}
\affiliation{\vspace{0.5em}QuTech, Delft University of Technology, 2628 CJ Delft, The Netherlands}
\affiliation{\vspace{0.5em}Netherlands Organisation for Applied Scientific Research (TNO), P.O. Box 155, 2600 AD Delft, The Netherlands}
\author{E.J. van Zwet}
\affiliation{\vspace{0.5em}QuTech, Delft University of Technology, 2628 CJ Delft, The Netherlands}
\affiliation{\vspace{0.5em}Netherlands Organisation for Applied Scientific Research (TNO), P.O. Box 155, 2600 AD Delft, The Netherlands}
\author{R. Hanson}
\email{Correspondence to: R.Hanson@tudelft.nl}
\affiliation{\vspace{0.5em}QuTech, Delft University of Technology, 2628 CJ Delft, The Netherlands}
\affiliation{\vspace{0.5em}Kavli Institute of Nanoscience, Delft University of Technology, 2628 CJ Delft, The Netherlands}

\date{\today}
             
\begin{abstract}
Entanglement distribution over quantum networks has the promise of realizing fundamentally new technologies. Entanglement between separated quantum processing nodes has been achieved on several experimental platforms in the past decade. To move towards metropolitan-scale quantum network test beds, the creation and transmission of indistinguishable single photons over existing telecom infrastructure is key. Here we report the interference of photons emitted by remote, spectrally detuned NV center-based network nodes, using quantum frequency conversion to the telecom L-band. We find a visibility of $0.79 \pm 0.03$ and an indistinguishability between converted NV photons around $0.9$ over the full range of the emission duration, confirming the removal of the spectral information present. Our approach implements fully separated and independent control over the nodes, time-multiplexing of control and quantum signals, and active feedback to stabilize the output frequency. Our results demonstrate a working principle that can be readily employed on other platforms and shows a clear path towards generating metropolitan scale, solid-state entanglement over deployed telecom fibers.
\end{abstract}

\maketitle
A future quantum internet~\cite{kimble_quantum_2008, wehner_quantum_2018}, built using quantum processor nodes connected via optical channels, promises applications such as secure communication, distributed quantum computation, and enhanced sensing~\cite{ekert_ultimate_2014, jiang_distributed_2007,gottesman_longer-baseline_2012}. In recent years, the generation of entanglement between remote processor nodes has been realized with ions and atoms~\cite{moehring_entanglement_2007, ritter_elementary_2012, hofmann_heralded_2012, stephenson_high-rate_2020}, Nitrogen-Vacancy (NV) centers in diamond~\cite{bernien_heralded_2013, humphreys_deterministic_2018}, and quantum dots~\cite{delteil_generation_2016, stockill_phase-tuned_2017}. Moreover, other platforms such as rare-earth doped crystals~\cite{Usmani2012,puigibert_entanglementmemories_2020, Lago-Rivera_telecommemories_2021}, atom-cloud based memories~\cite{Chou_measurement-basedentanglement_2005, Yu_telecommemoriesdistant_2020}, and mechanical resonators~\cite{Riedinger_mechanical-nonclassical-correlations_2016} have been used to explore distributed entangled states. 

Central to commonly used remote entanglement generation protocols~\cite{moehring_entanglement_2007, hofmann_heralded_2012, stephenson_high-rate_2020,bernien_heralded_2013, humphreys_deterministic_2018, delteil_generation_2016, stockill_phase-tuned_2017} is the propagation and interference of single photons that are entangled with stationary qubits in the nodes.
Scaling these schemes to many nodes and to long distances poses two main challenges. First, any source of distinguishability between the emitted photons needs to be removed to generate high-fidelity entangled states. Especially for solid-state emitters, this requirement is difficult to meet for a large number of nodes due to variations in the local environment of the emitters. Second, for long-distance connections photon loss in fibers is a dominant factor determining the rate at which the entanglement generation succeeds. Leading platforms for realizing processor nodes~\cite{Hucul_modularentanglement_2015, stephenson_high-rate_2020,Krutyanskiy_ionphotonlongrange_2019,Daiss_distantquantumlogic_2021,pompili_multinodenetwork_2021,vanleent_entanglingtelecom_2021} in a future quantum network have natural emission frequencies in the visible spectrum; fiber losses at these frequencies hinder scaling beyond a few kilometers.

\begin{figure*}
\includegraphics[scale=0.75]{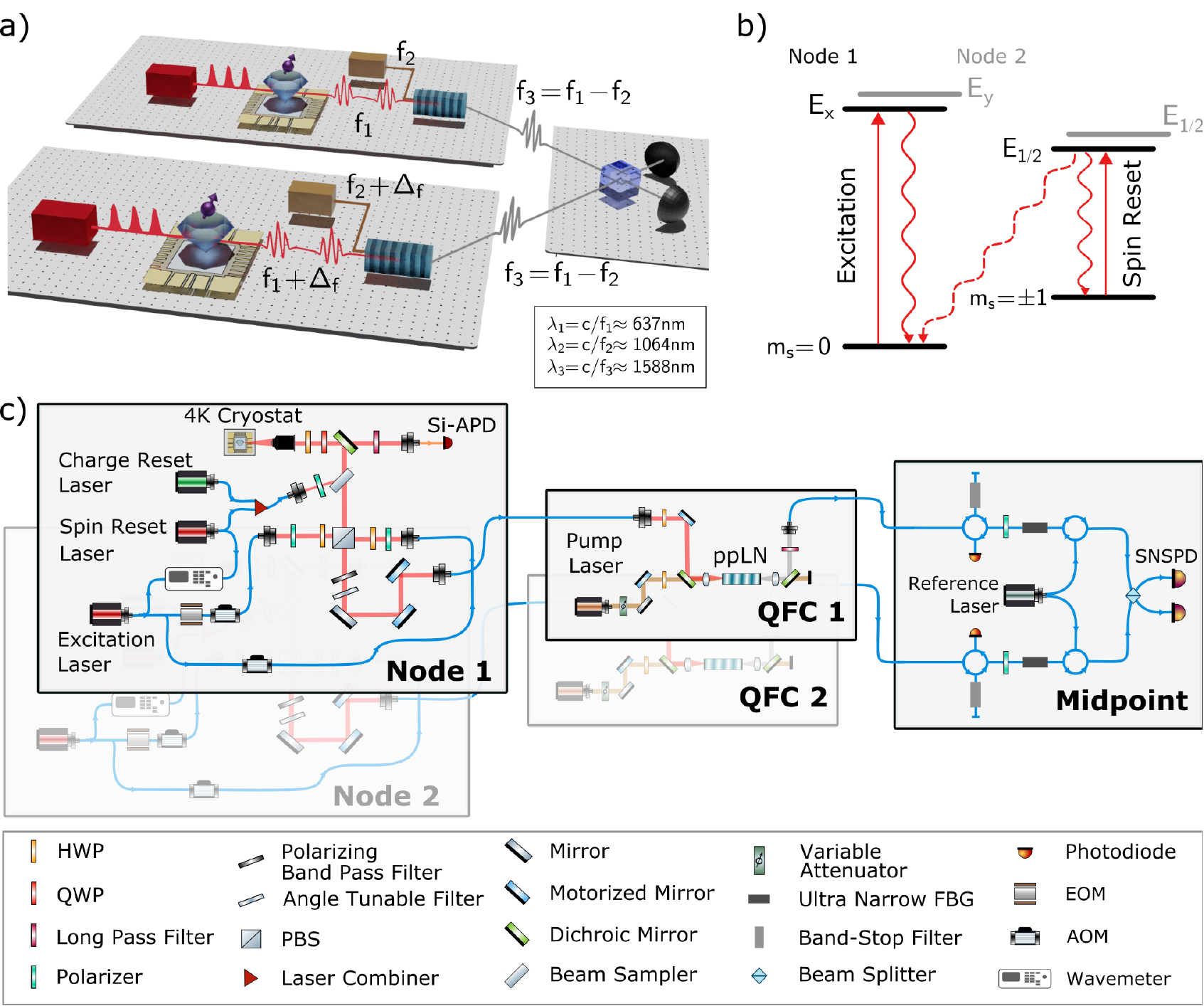}
\caption{\label{fig:Figure1}\textbf{Lay-out of the two independent network nodes and the midpoint.} a) Schematic of the main components of the set-up. Pulses excite the NV center, which emits single photons through spontaneous decay. The photons are converted to telecom wavelength and guided towards a midpoint placed in the neighboring lab, where they interfere on a beamsplitter~\ref{sec:supp_setup_details}. b) NV center level structure showing the spin levels in the ground state and the relevant optical transitions. The optical transitions of the two nodes have different energies due to local variations in strain (see text). c) Detailed schematic showing the optics used for full operation of the system. The nodes are identical, and no hardware is shared between the systems. Each node has a set of charge reset, spin reset and excitation laser, which are modulated, combined and focused via a high NA objective onto the NV center. The Phonon side-band (PSB) (Zero-phonon line (ZPL)) emission from the NV center is filtered using frequency (polarization) filtering and coupled into a multi (single-) mode fiber. The PSB emission is measured locally on an APD, while the ZPL emission is sent to the QFC module. Stabilization light is split off from the excitation path, and brought into the single-photon path via a polarizing beamsplitter. The QFC module contains remotely controllable optics to align the input, pump and converted fields to the waveguide on the ppLN crystal that converts both the single-photons and stabilization light. Polarization maintaining (PM) fibers transport the photons to a midpoint where the single photons are filtered and separated from the stabilization light using a combination of in-fibre filters. The single photons interfere on a beamsplitter and are measured by SNSPDs, while the stabilization light interferes with the reference laser and is measured on a photodiode.}
\end{figure*}

In this work, we show that both challenges can be addressed simultaneously by converting the coherent single-photon emission from NV centers (\SI{637}{\nano\metre}) to a single target wavelength in the Telecom L-band (\si{1565}-\SI{1625}{\nano\meter}) (Fig.~\ref{fig:Figure1}a). Using the pump lasers to compensate for local detuning and using active stabilization of the frequency of the converted field, we are able to decouple the natural emission wavelength of the emitters from the wavelength used for propagation and interference and build fully independent, modular quantum nodes. We demonstrate that this method enables the removal of spectral offsets over a broad frequency range ($>$\SI{3}{\giga\hertz}). Moreover, the chosen interference wavelength has low propagation losses over commercially available optical fibers, making it suitable for long-range single-photon transmission.

We validate our approach by measuring quantum interference~\cite{HOM_interference_1987} between telecom photons that are frequency converted from the emission of two remote NV-centers that are detuned by more than 100 linewidths. By comparing the data to a detailed model we extract both the major noise sources and the underlying indistinguishability of the converted NV photons.

\captionsetup[figure]{justification=raggedright}

\section{Independent quantum network nodes}
 We employ two independently operated quantum network nodes separated by a few meters on different optical tables. The nodes are connected to a midpoint located in two separate 19-inch racks. The relevant elements are depicted in Fig.~\ref{fig:Figure1}c. Each node operates a single diamond NV center as stationary qubit, hosted in a closed-cycle cryostat at $T\approx4K$. The relevant energy levels and optical transitions of the NV qubit are depicted in Fig.~\ref{fig:Figure1}b. The spin reset transition is used for spin initialization into $m_{s}=0$ (fidelity $>0.99$). We use the transition to the $E_{x/y}$ excited state to generate single photons: a coherent optical $\pi$-pulse ($\approx$\SI{2}{\nano\second}) brings the NV center to the excited state, followed by spontaneous emission (lifetime $\approx$\SI{12}{\nano\second}~\cite{Goldman_NVlifetime_2015,Kalb_NVlifetime_2018}). Both setups employ their own lasers and optical components that deliver and collect light to the NV center.

The nodes are equipped with a (nominally identical) Quantum Frequency Conversion (QFC) module. Here the light at \SI{637}{\nano\meter} is converted to \SI{1588}{\nano\meter} via a single-step Difference Frequency Generation (DFG) process. This process has previously been shown to preserve entanglement between the photon and an NV center qubit~\cite{Tchebotareva_TelecomSpinPhotonEntanglement_2019}. The QFC modules are based on waveguides in a periodically poled Lithium Niobate crystal (ppLN), where the large non-linear coefficient facilitates conversion of single photons using a strong \SI{1064}{\nano\meter} pump field. Various remotely controllable components allow for the remote and automated optimization of the QFC modules. While this single-step conversion has the upside of using a prevalent commercially available pump laser, it has the challenge of introducing spontaneous parametric down-conversion noise at the target wavelength by the pump laser due to imperfections in the crystal domains~\cite{Pelc_parametricnoisespdc_2010}.

After free-space filtering to remove the bright pump light, the frequency-converted light is guided to a central midpoint. In the midpoint, a series of filters separate photons at the target wavelength from stabilization light and filter out noise photons. To achieve a high suppression of any broad-band background light, we use a two-step filtering process. First, we use a reflection off a narrow Band-Stop filter of \SI{0.35}{\nano\meter}, followed by transmission through an ultra-narrow (FWHM of $\sim$\SI{50}{\mega\hertz}) Fiber Bragg Grating (abbreviated with UNF). The filters are connected via circulators, and an in-fiber polarizer is used to ensure optimal performance of the UNF (Fig.~\ref{fig:Figure1}B). After filtering, we interfere the two paths using a $50:50$ beam splitter, and detect the photons using single-photon superconducting nanowire detectors (SNSPDs).

\section{Realizing spectral stability via feedback}
For solid-state emitters, the local environment can influence the emission properties directly via electric, magnetic or strain fields. For instance, for NV centers in nominally low-strain samples the local strain environment can shift the emission frequency by more than 1000 times the linewidth~\cite{Ruf_membrames_2019}. While  direct tuning of emission wavelength is in principle possible on several platforms, e.g. by  strain~\cite{Meesala_strainengineering_2018} or via static electric fields~\cite{bernien_heralded_2013,stockill_phase-tuned_2017}, the range of tunability is in general limited. Also, direct tuning brings additional complexities in the device fabrication and can add significant experimental overhead. 

\begin{figure*}
\centering
\hspace*{-1cm}
    \includegraphics[scale = 1]{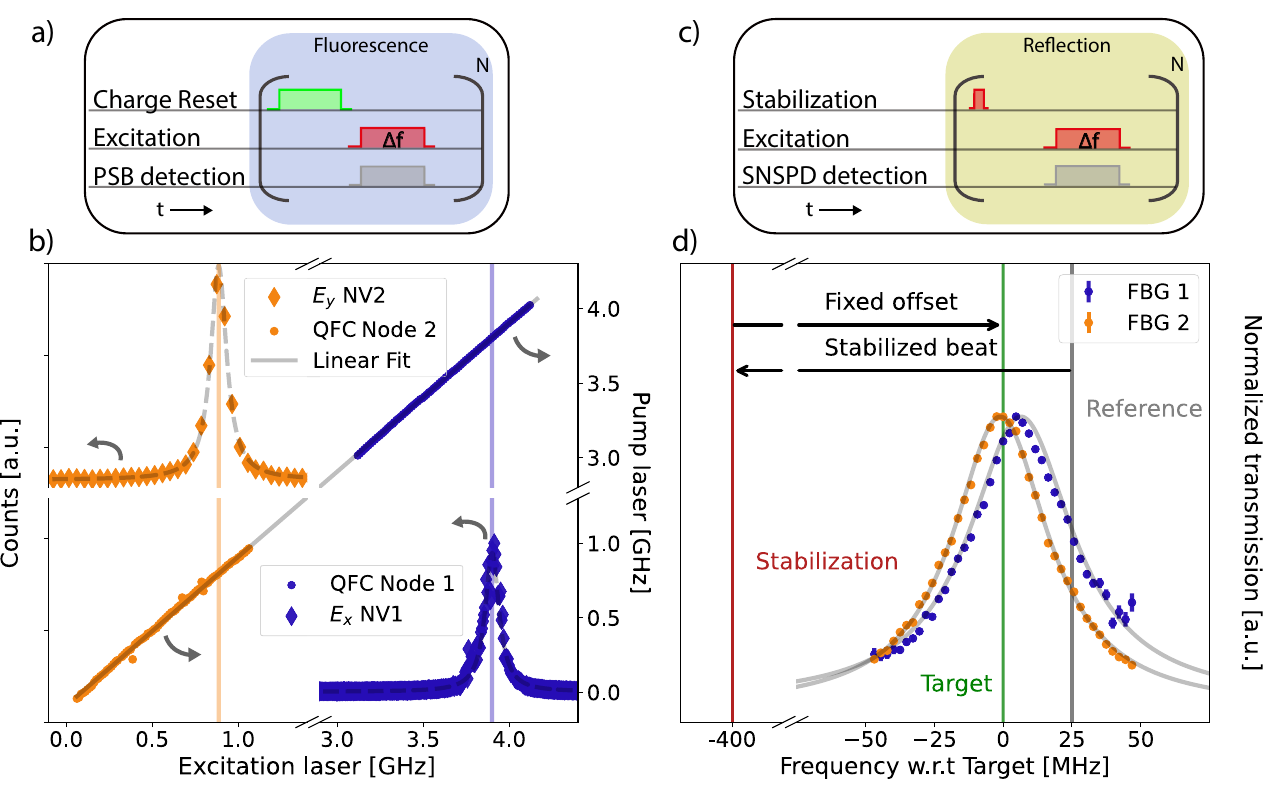}
    \caption{\label{fig:Figure2}\textbf{Removing spectral offset using conversion.} a)  Fluorescence measurement sequence used in b). A charge reset pulse is followed by repeated resonant excitation, during which fluorescence in the PSB is monitored. b) Resonant excitation spectra at Node 1 (Node 2) shown in blue (orange) diamonds, revealing the frequencies of the optical transitions (vertical lines) used for photon generation. The horizontal axis shows the optical excitation frequency with respect to a \SI{470477}{\giga\hertz} offset. Grey dotted lines are Lorentzian fits. The  right vertical axis shows the frequency of the QFC pump with respect to a \SI{281635}{\giga\hertz} offset. Blue and orange dots show the QFC pump laser frequency when the stabilization is active. Solid grey line is linear fit.  c) Pulse schematic for measuring the transmission through the ultra-narrow filter (UNF) shown in d). Reflections from the excitation light off the sample surface, which follow the same path as the resonant NV emission, are converted to the target telecom frequency and measured on the SNSPDs. The excitation frequency is swept by modulating the AOM frequency (see Fig.~\ref{fig:Figure1}c). Transmission data is corrected for the frequency dependence of the losses in the AOMs. d) Layout of relevant laser frequencies and transmission data of the ultra-narrow filters (UNF). The UNF transmissions (blue and orange dots) are actively stabilized via their temperature to half transmission of the reference laser (grey vertical line). The stabilization light (red vertical line) is detuned from the target (green vertical line) and therefore reflected off the UNF (see main text).}
\end{figure*}

We show that we can use the QFC process to remove the spectral offset between the NV emission without the need for direct tuning of the optical transition. An analogous scheme has recently been used on spectrally distinct quantum dots~\cite{Weber_TPQI29percent_2019, you_quantum_interference_dots_telecom_2021}; a key difference is that the NV center host diamond also contains a long-lived matter qubit and can function as a processing node in a quantum network~\cite{pompili_multinodenetwork_2021}. Using resonant excitation spectroscopy we find the optical transition frequency used for single-photon generation at each node (Fig.~\ref{fig:Figure2}a). We observe that these transition frequencies are separated by approximately \SI{3}{\giga\hertz} (about 100 natural linewidths), as shown in Fig.~\ref{fig:Figure2}B.

\begin{figure*}[t]
\centering
\includegraphics[scale = 0.66]{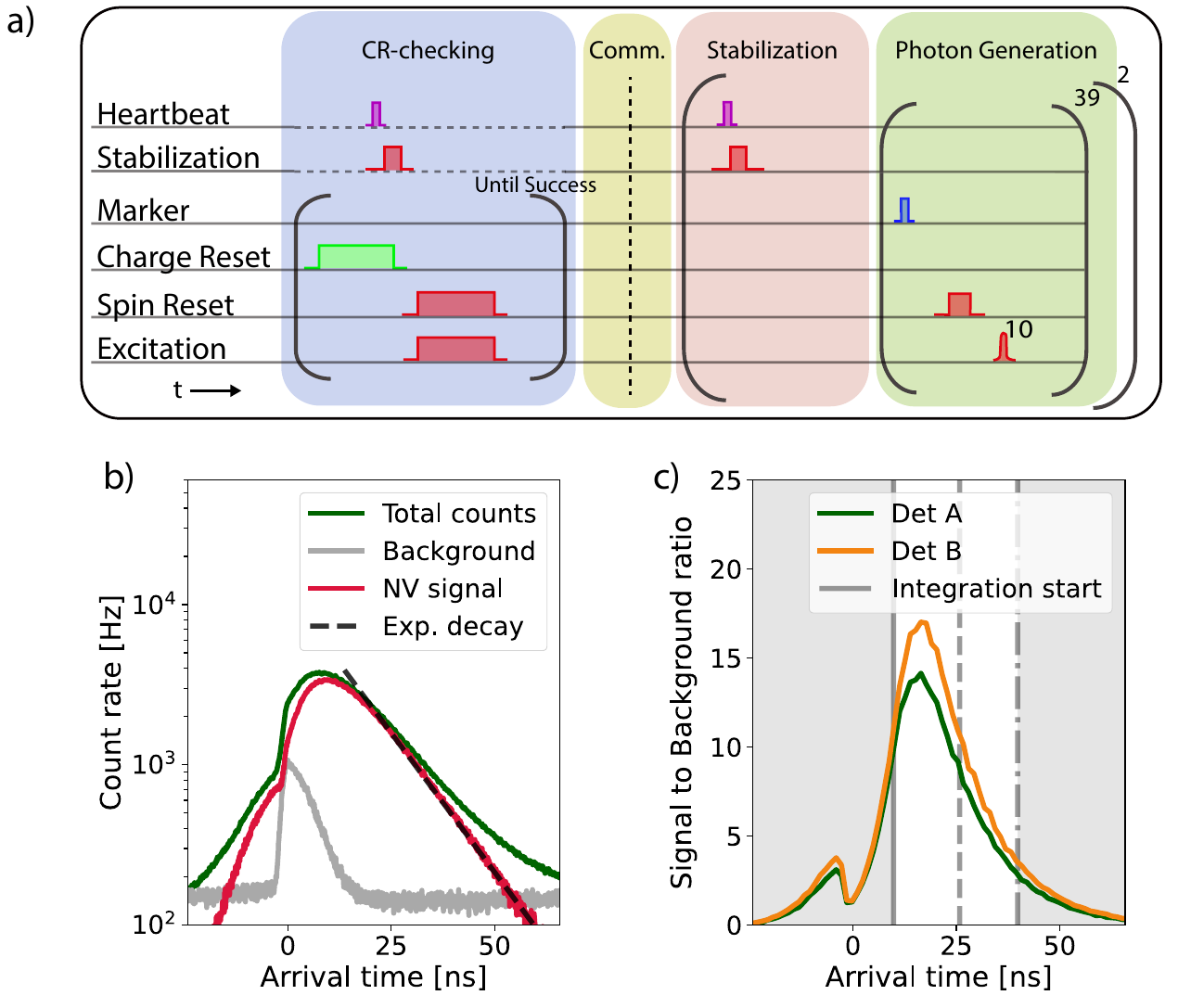}
\caption{\label{fig:Figure3} \textbf{Generating single-photons using a fixed heartbeat.} (a) Measurement sequence for synchronized generation of single-photons using a distributed heartbeat. For detailed timing see Supplementary Info~\ref{fig:supp_experimental_sequence}~\cite{supplementaryinfo}. (b) Histogram of SNSPD counts in a single excitation round, averaged over $\approx$ \SI{2.8e11}{\ }repetitions and analyzed in \SI{80}{\pico\second} bins. The measured count rate (green) is a combination of background (grey) and NV fluorescence (red), which is calculated by subtracting the background from the total counts. The dotted line depicts exponential decay with \SI{12.5}{\nano\second} lifetime and serves as a guide to the eye. The NV signal before time $0$, the peak of the excitation pulse, is due to the non-perfect extinction ratio of the devices generating the optical excitation pulse. The background data was taken continuously over 24 hours to include any drifts that occur over the same timescale as the signal data. (c) Signal-to-background ratio for both detectors as calculated from data in (b) (solid lines). The solid vertical grey line shows the start of our chosen detection window, whereas the dashed (dash-dotted) line shows the end for the data shown in Fig.~\ref{fig:Figure4}a (Fig.~\ref{fig:Figure4}b). The difference between the two curves is due to non-equal detector performances.}
\end{figure*}

To bring the NV photons to the same target frequency, we realize a scheme that locks the pump laser at each node to the frequency difference between the excitation laser at that node (and hence the NV emission frequency) and a joint Telecom reference laser at the midpoint. To achieve this lock, we use a split-off of the excitation laser, offset in frequency by a fixed \SI{400}{\mega\hertz}, as stabilization light (see Fig.~\ref{fig:Figure2}D). We propagate this stabilization light through the same frequency-conversion path as the NV photons. Due to the frequency offset, the stabilization light is reflected at the UNF, travelling backwards towards the first circulator where it exits (see Fig.~\ref{fig:Figure1}c). Light from the joint reference laser is inserted at the second circulator, from where it propagates in opposite direction through the transmission flank of the UNF, also exiting on the first circulator. Here, the interference with the stabilization light is measured on a photo diode yielding the error signal for the lock (see Supplementary Info~\ref{fig:supp_freq_lock}~\cite{supplementaryinfo}). We close the loop by applying feedback to the pump laser, imprinting the same frequency shift on both the single photons and stabilization light.
By transmitting the reference laser \SI{25}{\mega\hertz} detuned from the transmission peak of the second ultra narrow filter, the DC amplitude on the same photo diode serves as an error signal for active temperature stabilization of the UNF. Typical transmission profiles of the temperature-stabilized UNFs and the respective light-field frequencies are shown in Fig.~\ref{fig:Figure2}d. The small deviation from ideal peak transmission at the target frequency is due to unaccounted background voltage of the photo diode and the slight difference in FWHM of the two UNFs. The remaining thermal drifts of about \SI{1}{\mega\hertz} have only a minor ($\sim$1$\%$) effect on the transmission (see Supplementary Information~\ref{sec:supp_UNF_stability,fig:supp_UNF_transmission}~\cite{supplementaryinfo}). Note that the part of the reference light that is reflected off the ultra-narrow filter exits the circulator towards the SNSPDs and thus needs to be taken into account when designing the experimental sequence (see next section).

We verify our frequency locking by sweeping the excitation laser and monitoring the resulting pump laser frequency in Fig.~\ref{fig:Figure2}b. We observe the expected linear relationship: a change in excitation laser frequency is precisely compensated by the pump laser frequency to always yield the same target frequency across the full tuning range. A linear fit yields the target frequency of \SI[separate-uncertainty]{1587.5298 \pm 0.0001}{\nano\metre}. This data demonstrates the ability of our frequency-locked down-conversion system to robustly compensate for a wide range of detunings.

\section{Photon generation at the target telecom wavelength}
We now turn to the generation of single telecom photons by the nodes as used for the measurement of two-photon quantum interference at the midpoint. The measurement sequence involves four stages (see Fig.~\ref{fig:Figure3}), which are synchronized across the nodes by a fixed electronic ``heartbeat'' every \SI{200}{\micro\second}. This heartbeat is derived from a GPS-disciplined atomic clock positioned in the midpoint, which is distributed over telecom fibers via the White Rabbit Precision Time Protocol~\cite{Dierikx_whiterabbitprotocol_2016}. The first \SI{2.5}{\micro\second} following each heartbeat are used for the error signal generation for the frequency lock. This scheme allows the frequency lock to operate without knowledge of the state of the nodes, which reduces the complexity and rounds of communication needed. Moreover, it enables the autonomous operation of each of the nodes, using their own independent hardware to control the NV center and generate single photons.  

In the first stage of the measurement sequence, a Charge-Resonance (CR) check is performed at each node to ensure that the NV centers are in the correct charge state and their transitions are on resonance with their respective spin reset and excitation lasers~\cite{Robledo_SSRO_2011, bernien_heralded_2013}. In case a CR check fails, a charge reset laser pulse is applied and a new CR check is started; this protocol is repeated until success. Importantly, the CR check can be run in parallel with the frequency locking as the stabilization light for the lock does not reach the NV center nor the local PSB detectors; hence the CR checks can run independently of the heartbeat. The second phase starts once the CR check is passed on a node, where a digital trigger from the micro-controller signals the readiness to the other node. After the readiness of both nodes has been communicated, the heartbeat at which they move to the third stage is agreed upon (Fig.~\ref{fig:Figure3}a and Supplementary Info~\ref{fig:supp_experimental_sequence}~\cite{supplementaryinfo} for timings and more information).

The third stage of the measurement sequence is used for the time-multiplexed frequency stabilization, as described above. In the fourth stage, we repeatedly apply a block consisting of a spin reset pulse (\SI{1.5}{\micro\second}) followed by 10 optical $\pi$-pulses, ideally generating a train of 10 NV photons. A time-tagged digital signal marks the times at which the photon generation takes place.  This block is repeated 39 times per heartbeat period. After two heartbeat periods, the system returns to the first stage (CR checks). Note that during the third and fourth stage, time-multiplexing the operations on the NV center and the error signal generation for the frequency lock is critical as the stabilization light and the reference light both leak into the single-photon detection path.

\begin{figure*}[t!]
\centering
\includegraphics[scale = 0.66]{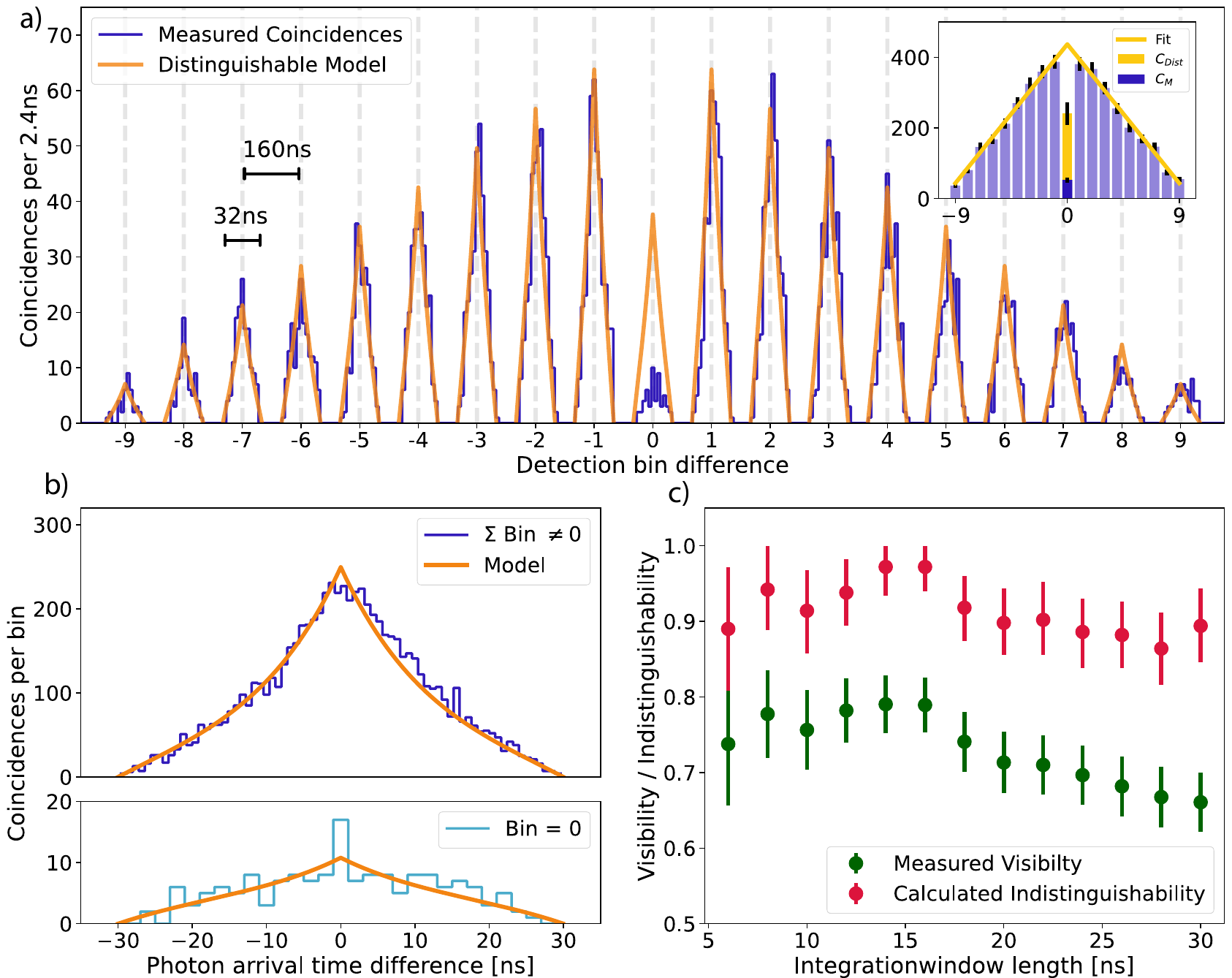}
\caption{\label{fig:Figure4}\textbf{Two-photon quantum interference at telecom wavelength.} a) Histogram of measured coincidences (blue) for the analyzed time-bins, overlaid with a model assuming distinguishable photons (orange), based on independently determined parameters. Histogram binsize is \SI{2.4}{\nano\second}. Vertical lines depict the time-origin of each detection bin difference. Horizontal scale bars show the relevant timescales in and between the detection bin difference. Inset: extraction of the interference visibility, corrected for the imbalance of measured photons per excitation of the two nodes, using a linear fit to the total counts per bin difference (see text).  b) Histogram showing temporal shape of the non-zero (zero) bin difference coincidences shown at the top (bottom) panel. We overlay the data with the same model (see Supplementary Info), taking into account brightness and background rates and indistinguishability of converted NV photons of 0.9 (see c)). Binsize for top (bottom) panel is \SI{0.8}{\nano\second} (\SI{2}{\nano\second}) c) Measured visibility and extracted indistinguishability of converted NV photons for varying window length. Errorbars for visibility/indistinguishability are $1\sigma$/$68\%$ confidence interval. Green (blue) circled points indicate data corresponding to the window length shown in Fig.~\ref{fig:Figure4}a (Fig.~\ref{fig:Figure4}b).}
\end{figure*}

We analyze the resulting telecom photon detection rate in Fig.~\ref{fig:Figure3}b (green line). We show the events observed in a single detector, aggregated over all single excitation rounds and both nodes. We denote $t = 0$ as the relative time of excitation.  A sharp increase in count rate is observed when the $\approx$\SI{2}{\nano\second}-wide optical $\pi$-pulse starts, followed by a slower decay dominated by spontaneous emission of the NV centers.

A data set displaying only the noise counts and the counts due to leakage of the excitation $\pi$-pulse (grey line) is independently generated by detuning the excitation laser by \SI{1}{\giga\hertz}. The observed uniform background consists of intrinsic detector darkcounts (\SI{5}{\hertz} per detector), counts induced by detector blinding from leaked reference and stabilization light (\SI{35}{\hertz} per detector), and SPDC photons from the QFCs ($\approx$\SI{150}{\hertz} per detector). The leakage of the excitation $\pi$-pulse reflected off the sample is clearly visible. By subtracting this background from the data, we isolate the frequency-converted NV signal (red) displaying the characteristic exponential decay.

In remote entanglement experiments, the effect of noise counts can be mitigated by defining a heralding detection time window: only photon counts in this window are taken as valid entanglement heralding events~\cite{bernien_heralded_2013}. In general, setting the heralding window involves a trade-off between high signal-to-background and thus high fidelity (favouring shorter windows) and success rate (favouring longer windows). In Fig.~\ref{fig:Figure3}c we plot the signal-to-background ratio (SBR) for the two detectors as a function of photon detection time. The SBR is bounded on one side by the leaked excitation pulse and on the other side by the NV signal approaching the uniform background. For the analysis of the two-photon interference visibility (Fig.~\ref{fig:Figure4}a), we apply a detection window in which the average SBR exceeds 10 (Fig.~\ref{fig:Figure3}c, up to dashed line). For a more detailed comparison of our model to the data (Fig.~\ref{fig:Figure4}b) we use an extended window (up to dash-dot line). In order to maintain the same SBR throughout the experiment, we employ a system of automatic optimization based on the live monitoring and processing of the single photon detection events (see Supplementary Info~\ref{sec:supp_automatic_calibrations},~\ref{fig:supp_auto_calibrations} and~\ref{fig:supp_freqlock_calibrations}~\cite{supplementaryinfo}).

\section{Two-photon quantum interference}
Next we investigate the distinguishability of the photons emitted by the two nodes by analyzing their quantum interference. For two fully indistinguishable photons impinging on the input ports of a balanced beam splitter, quantum interference leads to vanishing probability to detect one photon in each output port~\cite{HOM_interference_1987}, while for fully distinguishable photons this probability is 0.5~\cite{Paul_interferencephotons_1986, Bouchard_TPQIoverview_2021}. From the (properly normalized) coincidence counts in the two detectors we can thus extract the distinguishability of the photons.

Figure~\ref{fig:Figure4}a) shows the measured coincident detections between the two output arms without any background subtraction. Each excitation round is treated as a ``detection bin'' in which a photon can arrive. We analyze the coincidences per block of 10 excitation pulses, defined as a click in both detectors in the same or two different detection bins. This leads to a maximum detection absolute bin difference of 9 and a coincidence probability increasing linearly towards 0 bin difference. We overlay the data with a model based on independently determined parameters, treating the photons as completely distinguishable (see Supplementary Information~\cite{supplementaryinfo}). For the non-zero bin differences, in which the NV photons are fully distinguishable by their arrival time difference of at least 10x the lifetime, the model shows excellent agreement with the measured coincidences. In stark contrast,  we observe a strong reduction in measured coincidences compared to the model for the zero bin difference. This drop in coincidences when the photons arrive in the same bin is the hallmark of two-photon quantum interference and forms the main result of this work.

The observed visibility is defined as $V = 1 - \frac{C_{M}}{C_{Dist}}$, with $C_{M}$ the measured number of coincidences, and $C_{Dist}$ the coincidences we would have measured at zero bin difference in case the photons were completely distinguishable. In the inset of Fig.~\ref{fig:Figure4}a we show the method of extracting the visibility. First we use a linear fit to the total distinguishable coincidences per detection bin difference to get $C_{E}$, the extrapolated coincidences for $0$ bin difference. From this value we extract $C_{Dist}$, by correcting for the imbalanced emission rates (see Supplementary Inf~\ref{sec:supp_ModelofCoincidence}~\cite{supplementaryinfo}). The resulting visibility is $V=0.79 \pm 0.03$, which is well above the classical bound of $0.5$~\cite{Paul_interferencephotons_1986, Bouchard_TPQIoverview_2021}, proving the successful demonstration of quantum interference of single photons in the telecom L-band.

A more detailed picture of the temporal shape of the coincidences allows us to test our model with more precision (Fig.~\ref{fig:Figure4}b). The accumulated coincidences for non-zero bin difference (top panel) show a characteristic shape dominated by the exponential decay of the NV emission. The data is well described by our model that takes into account the temporal shape of the NV-NV, background-NV and background-background coincidence contributions (derivation in Supplementary Info~\ref{sec:supp_ModelofTemporal}~\cite{supplementaryinfo}). The temporal histogram of coincidences within the same bin (bottom panel) shows a good match with the temporal shape predicted by our model. In particular, we observe no reduction of coincidences at $0$ time delay, consistent with the visibility being limited by background counts rather than frequency differences between emitted photons~\cite{Legero_TPQI_2003,Kambs_limitationsTPQI_2018}.

With our knowledge of the background and signal rates we can extract the degree of indistinguishability of the emission coming from our NV centers. We perform a Monte-Carlo simulation of our dataset using the independently determined parameters and apply Bayesian inference to find the most likely value of the indistinguishability, given our measured result (see Supplementary Info~\ref{sec:supp_bayesian_inference} and~\ref{fig:supp_pdf_indist}~\cite{supplementaryinfo}). 

In Fig.~\ref{fig:Figure4}c we plot the visibility and the extracted photon indistinguishability for increasing detection time window lengths. While the visibility drops for longer windows consistent with the decreasing signal-to-background ratio, the indistinguishability of the NV photons remains high around 0.9. We note that this latter value is similar to values found for NV-NV two-photon quantum interference without frequency conversion~\cite{humphreys_deterministic_2018}, confirming that our conversion scheme including the frequency stabilization to a single target wavelength preserves the original photon indistinguishability, and enables solid-state entanglement generation via entanglement swapping.

\section{Conclusion and Outlook}
We have shown quantum interference of single photons emitted by spectrally distinct NV centers, by converting them to the same telecom wavelength. We have demonstrated an actively stabilized Quantum Frequency Conversion scheme using Difference Frequency Generation on fully independent nodes. The design and implementation allow for the scheme to be used at large distances. Furthermore, the techniques can be readily transferred to other quantum emitters in the visible regime with minimal adaptations to the conversion optics and control schemes used.

Future improvements to our system can increase the performance in multiple ways. First, adapting our optical design to prevent detector blinding by the stabilization light can lower the detector contribution to the background counts to the design level of \SI{5}{\hertz}. Second, a different approach to the QFC technique based on a bulk crystal may remove the (currently dominating) SPDC background noise due to poling irregularities. Third, the signal level of collected coherent photons from the NV centers could be improved significantly by use of an open microcavity~\cite{riedel_enhancementphotongeneration_2017, ruf_purcellenhanced_2021}. In particular, achieving a fraction of coherent emission of 46\% as reported in Ref.~\cite{riedel_enhancementphotongeneration_2017} would raise the signal-to-background ratio above 200. Finally, by extending the hardware we can stabilize the optical phase of the single photons emitted by the NV centers, enabling entanglement generation upon heralding of a single photon~\cite{pompili_multinodenetwork_2021}.

By combining the protocols demonstrated here with established spin-photon entangling operations and photon heralding at the midpoint~\cite{bernien_heralded_2013, humphreys_deterministic_2018}, remote NV centers can be projected in an entangled state via telecom photons. Owing to the low propagation loss of these photons and extendable control scheme, our results pave the way for entanglement between solid-state qubits over deployed fiber at metropolitan scale.

\begin{acknowledgments}
We would like to acknowledge Matthew Weaver and Matteo Pompili for fruitful discussion during the model development and data analysis, Ludo Visser and Pieter Botma for software support during the measurements. We thank Martin Eschen and Boudewijn van den Bosch for their contributions to building and maintaining parts of the setup, and Wouter Koek for simulations aiding the design. We thank Klaas Jan de Kraker, Sander Kossen, and Emanuele Uccelli for managing many tasks in the project/sample production. We thank Ruud Schmits and Jacob Dalle for their contributions to the sample production process development.

We acknowledge funding from the Dutch Research Council (NWO) through the project “QuTech Part II Applied-oriented research” (project number 601.QT.001), VICI grant (project no. 680-47-624) and the Zwaartekracht program Quantum Software Consortium (project no. 024.003.037/3368). We further acknowledge funds from the Dutch Ministry of Economic Affairs and Climate Policy (EZK), as part of the Quantum Delta NL programme, and Holland High Tech through the TKI HTSM (20.0052 PPS) funds.

\textbf{Author contributions}
A.S., K.L.vd.E, E.v.Z and R.Hanson devised the experiment. J.M., A.V, R.Hagen, A.K., B.B. designed and built the experimental hardware and the frequency conversion modules. S.C., J.v.R and I.t.R developed the measurement-, timing and control software framework. M.M. and A.M.E. grew and prepared the diamond device substrates. R.V., A.S. and J.M. characterized and fabricated the diamond devices. A.S., K.L.vd.E, M.C.R. and S.O.F carried out the experiments and collected the data. A.T., A.S. and M.C.R developed supporting simulations. A.S., K.L.vd.E and M.C.R analyzed the data. A.S. and R.Hanson wrote the main manuscript with input from K.L.vd.E and M.C.R. A.S., K.L.vd.E and M.C.R wrote the supplementary information. All authors commented on the manuscript. J.M. and E.v.Z coordinated the experimental work. R.Hanson supervised the quantum experiments.

\end{acknowledgments}

\bibliography{main}
\bibliographystyle{my_plain}
\onecolumngrid

\clearpage
\begin{center}
\textbf{\large Supplementary Materials}
\end{center}
\makeatletter
\renewcommand{\theequation}{S\arabic{equation}}
\renewcommand{\thefigure}{S\arabic{figure}}
\renewcommand{\thetable}{S\arabic{table}}
\renewcommand{\thesection}{S-\Roman{section}}
\makeatother
\setcounter{equation}{0}
\setcounter{figure}{0}
\setcounter{table}{0}
\setcounter{section}{0}
\section{Optical set-up}
\label{sec:supp_setup_details}
Our experiments are performed using two nominally identical quantum network nodes. Each node houses a Nitrogen-Vacancy (NV) center in a high-purity type-IIa chemical-vapor-deposition diamond cut along the $\langle 111 \rangle$  crystal orientation (Element Six).
Both samples have a natural abundance of carbon isotopes. 
Fabrication of solid immersion lenses and an anti-reflection coating on the diamond samples enhances the photon-collection efficiencies from the NV centers. The ground-state spin levels are split using a small permanent magnetic field aligned with the NV axis of $\approx$ \SI{30}{G}.

Experimental equipment used for each node is summarized in  Tab.~\ref{tab:supp_equipment}. Node 1 and 2 are in the same laboratory, around \SI{7}{m} apart. The optical fibers that connect Node 1 and 2 with the first midpoint $19$" rack containing the filters are PM fibers of \SI{3}{m} and \SI{10}{m} long, respectively. After the filters, both are connected to the beamsplitter and subsequently the SNSPDs in a separate $19$" rack in the room next door with \SI{10}{m} long PM fibers.

Details of the optical lay-out of the free-space and in-fibre optics used are shown in figure 1B. For initialization in the correct charge and spin state, we use a combination of off-resonant (\SI{515}{\nano\metre}) and resonant (\SI{637}{\nano\metre})  excitation respectively. Both lasers are combined using in-fibre optics, and coupled into the free-space part using a beam sampler optimized for transmission. For single-photon generation, we use a second resonant laser tuned to a spin-preserving transition. Both resonant lasers are stabilized using a wave meter. The main part of the excitation light passes electro- and acoustic- optical modulation (EOM and AOM) for fast switching, and is coupled into the free-space path using a central polarizing beam splitter (Thorlabs). Combined, this light is guided to the high NA microscope objective (Olympus 100x 0.9NA) using a dichroic mirror (DM) (Semrock), after which the polarization is set at an optimum for cross-polarization.
The second part of the excitation laser is sent to a separate set of AOMs that provide an offset in frequency to form the stabilization light and coupled in on the opposite side of the central PBS. The main purpose of this light is to provide a fixed frequency reference of the single photons generated by the excitation light, and its purpose can be extended to stabilize the phase of the relevant phases for entanglement generation \cite{pompili_multinodenetwork_2021, quantumnetwork_patent}. 

Emission from the NV-centre propagates backwards through the set-up, where the phonon side-band (PSB) is separated from the main path using the DM, filtered by additional long-pass filters (Semrock) and coupled into a multi-node fibre to be detected locally by an APD (LaserComponents). During the measurement, detection in this path is used for live monitoring and diagnostic purposes. The resonantly emitted photons in the zero-phonon line (ZPL) are guided to the central PBS, separated from the excitation light based on polarization (Thorlabs), and filtered using a band pass filter (Semrock). We use a deformable mirror and a set of motorized mirrors to couple in both the single photons and frequency control light simultaneously into the same PM SMF. Using these components, we can periodically optimize the coupling remotely, greatly improving the long-term performance and remote operation capabilities. By coupling into a fibre, we can decouple the alignment of the free-space optics from the QFC optics, simplifying the set-up and allowing for easy exchange of different QFC modules (full details of equipment list in Tab.~\ref{tab:supp_equipment} and \ref{tab:supp_equipment_qfc}.

\section{Overview of timing electronics}
Interfering photons emitted by two NV centres with high fidelity places strict requirements on the arrival time and thus the generation time of the single photons. Future long distances between our nodes prevent us from using a single waveform generator to meet these requirements, and a more scalable approach is needed, which we have employed for the current experiment. The lay-out is shown in figure \ref{fig:supp_overview_adapted}  (full equipment in \ref{tab:supp_equipment_timing}), showing how a single GPS disciplined clock is shared across the nodes via White-Rabbit enabled switches. The resulting synchronization pulses are then coherently split and distributed to the devices requiring timing synchronization. The microcontrollers EVENT cycles are externally triggered by a \SI{1}{\mega\hertz} pulse. The AWG sequencers are referenced externally by the \SI{10}{\mega\hertz}. Furthermore, the waveform generation inside the AWG is triggered using the \SI{5}{\kilo\hertz} heartbeat, after which the AWG sequencer can realize a sub nanosecond delay. Together, this allows for the long range synchronization of the experimental sequence with a jitter of \SI{100}{\pico\second}.

For future experiments, the heartbeat can be adjusted by changing the settings of the heartbeat generator. This would allow to run the frequency stabilization at a higher rate needed for the higher feedback bandwidth for phase-locking of two remote excitation lasers. 

\section{Overview experimental sequence.}
The experimental sequence is given in more detail in Fig.~\ref{fig:supp_experimental_sequence}. The CR-check procedure contains a bright off-resonant charge reset pulse to probabilistically re-ionize the NV centre into the negative charge state. We then verify that the lasers are resonant with the right transitions by monitoring the fluorescence during excitation with both the spin reset and excitation laser. The counts during this interval are compared to a threshold determined before the experiment. If past this threshold, we assume to be on resonance and in the right charge state. This procedure is repeated until success, and has a typical passing rate of $5-10\%$, taking on average \SI{1.5}{\milli\second}\cite{Robledo_SSRO_2011}. The amount of experimental sequences per CR-checking round is a balance between ionization to the $NV^{0}$ state during the sequence and data rate. At the start of each CR-check we find the NV centres to be ionized in about $10\%$ of the cases per node.

The communication between the nodes signalling the ready state is done by exchanging a digital trigger between the micro-controllers in charge of the CR-check process. By using the predetermined travel time of the communication, each micro-controller can calculate the next available starting time upon receiving the digital trigger of the opposite node. Using this information it triggers the AWG in advance of that heartbeat, reliably triggering the AWGs that switch to playing the waveforms that generate the optical pulses needed for photon generation. During the live analysis of the data, the experiment markers are used to signal the generation of single photons. These markers are in sync with the heartbeat present on the midpoint, and allow for the faithful recovery of the single photon events. Future improvements can be made by actively heralding the arrival of photons in the midpoint, conditioned on receiving the experiment signal.

\section{Data collection and processing}
\label{sec:supp_DataCollection}
The data taken shown in Fig.~\ref{fig:Figure3} and Fig.~\ref{fig:Figure4} was taken over the course of $17$ days. To allow for long-duration remote experiments, we operate the set-ups with minimal manual in-situ intervention, and employ a multitude of automatic calibrations. We determined the amount of data to collect by using previous measurements of coincidence rates and targeting an approximate statistical errorbar of $3\%$ on the measured visibility.

The data was collected in batches of $\approx24$ hours, during which the set-up was operating fully autonomously. Remote monitoring was made possible by live viewing in a Grafana dashboard showing performance and environmental data pushed to a database. Live warnings of parameters that went outside pre-defined ranges provided $24/7$ warnings, and the possibility to manually intervene remotely.  After each block of $24$ hours, manual routine inspections were done to check the sample and set-up status. As an example of the benefits of this system, the interference data was collected over a period of 17 days, of which $78\%$ of the wallclock time under autonomous operation of the complete system, during which no human operator actively controlled the experiment.

\subsection{Data processing}
We employ various data processing steps during the experiment. First, all the raw data generated by the Adwin counting modules describing the result of the CR-checking process on the nodes is written to disk every $\approx 2$ of minutes, and a next block of measurements is started. Simultaneously the stream of time tags of all synchronizations (heartbeat and experiment marker), and all SNSPD events outside of the blinding window, are tagged using \SI{80}{\pico\second} bins, and saved to disk on dedicated node PCs every couple of seconds, totalling to $\approx1.2$Tb of uncompressed data.

\begin{figure}[ht]
    \centering
	\includegraphics[scale = 0.5]{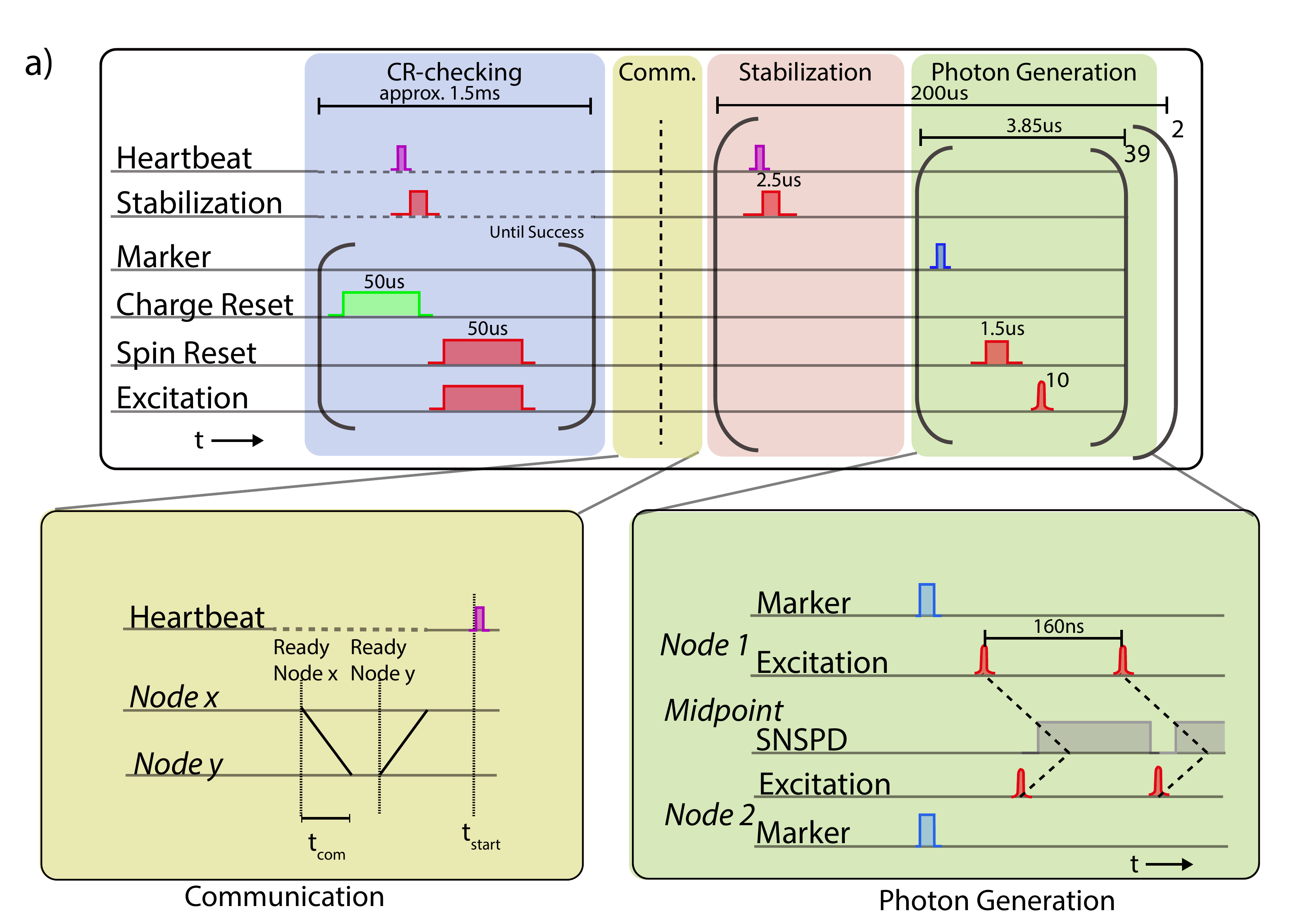}
	\caption{\label{fig:supp_experimental_sequence}. Schematic showing pulses played during the experiment. The timescale is indicated above the pulses and sequence stages. The communication step (bottom left) is symmetric whether node1 or node2 succeeds first in passing the CR check. The heartbeat where the next stage is started is shown. Using the Marker signals, the corresponding coincidence windows in the SNSPDs in the midpoint can be calculated.}
\end{figure}

Because performing the analysis on large amounts of raw data is challenging, we employ live processing that generates significantly smaller files. This processing stores only the photon events that might be of interest for further analysis, by looking at the time bins where converted single photons from the nodes are expected to arrive in the midpoint (see Fig.~\ref{fig:supp_experimental_sequence}). The data stored when an event occurs is given in Tab.~\ref{tab:supp_photon_filtering}, which can be extracted from the combined stream of timetagged events from all 3 timetaggers. One experiment marker can have more than one event (e.g. coincidences). The size of this dataset is less than $0.5\%$ of the raw data generated, making it much more manageable.

While analyzing the coincidence data we noticed that some blocks contain many events that happen at time intervals much shorter than the dead time of our detectors (\SI{20}{\nano\second}). We suspect that these events are due to (yet unexplained) resonances in the biasing electronics of the SNSPDs. To filter out these events we ignore blocks of measurements in which more than $2$ detection events have happened in a single \SI{160}{\nano\second} window for both the SNSPDs. We can give an upper bound on how many events we would expect of this kind based on our signal rate by taking the maximum singles detection rate in Fig.~\ref{fig:Figure3}b (\SI{3000}{\hertz}) for the full duration of the \SI{160}{\nano\second}. Then, the probability of having more than $2$ counts in both nanowires simultaneously is given by $P(C>2)^2$ with
\begin{equation}
   P(C>2) = 1 - P(C=0) - P(C=1) - P(C=2)
\end{equation}
the chance of having more than $2$ counts in one nanowire. For the amount of repetitions of our experiment ($282226000000 \approx 2.8*10^{11}$), we thus conservatively expect $0.015$ of these occurrences. When applying this filter we remove $\approx 36$ blocks which equates to $0.05\%$ of the total data. 

\section{Automated calibrations}
\label{sec:supp_automatic_calibrations}
Keeping multiple nodes at their pre-calibrated performance level in parallel requires an automated calibration framework, in which several setup-specific calibrations can be performed both in parallel and successively, where the order of calibrations is conditional on setup specific inter-dependencies. After a constant amount of optical excitations, each node's performance is evaluated against several parameters that are indicative for the NV photon emission, conversion and collection rate per node, listed below in Sec.~\ref{sec:supp_performance_parameters}. Every parameter has a specific threshold chosen to reflect the manually calibrated performance level at the time of initialization of each dataset taken per day. A visual representation of the automated calibration framework is shown in Fig.~\ref{fig:supp_auto_calibrations}. Violation of the pre-calibrated parameter thresholds triggers adding specific calibrations to the `calibration queue', where the calibrations are ordered by setup specific inter-dependencies and subsequently executed, parallelizing calibrations on nodes as much as possible.\\
\\
Every individual calibration that will block the optical path output of the QFC temporarily freezes the frequency lock's feedback loop for the duration of the calibration, schematically shown in Fig.~\ref{fig:supp_freqlock_calibrations}. We require this when calibrating the QFC efficiency with path-blocking flip-in power meters, as well as using a diagnostic flip-in mirror to measure NV excitation resonant laser light and photons coupling in to the QFC. The lock feedback loop is resumed after every such a calibration to ensure the locking feedback scheme can keep up with longer term drifts of the locking beat frequency.

\subsection{Evaluated performance parameters}
\label{sec:supp_performance_parameters}
\begin{itemize}
    \item \textbf{Count-per-shot of PSB photons ($\mathbf{CPS_{PSB}}$}): The amount of photon detections in the PSB per excitation attempt contains information on how well we address the relevant transition with the optical excitation pulse. A reduction in $\mathrm{CPS_{PSB}}$ can thus trigger all available node-related calibrations, including re-calibration of the power of the excitation pulse, but only of the specific node on which this is detected.
    \item \textbf{Average CR-check passing rate}: Indicative of the stability of optical addressability of the NV center's transitions.
    \item \textbf{Fixed amount of optical excitations}: Recalibration of each node is performed regardless of performance thresholds after a fixed amount of attempts, both for logging purposes of small drifts as well as certainty about bringing the setup to a well calibrated known state, mitigating any drifts we could not observe while the experiment is live.
    \item \textbf{Count-per-shot of telecom photons from NV excitation ($\mathbf{CPS_{tel}}$}): The amount of single telecom photons incident on the SNSPD detectors per excitation attempt show a performance of the entire node: the entire optical path the single photons traverse until detection at the SNSPD detectors. This also includes the intrinsic performance of the QFC, which is calibrated separately.
    
\end{itemize}

These measurements also enable us to collect the data crucial for the further analysis of the two-photon quantum interference: the probability of detecting a photon from Node $1$ or $2$ in detector $A$ and $B$ and the rate of background counts detected in detector $A$ and $B$. An overview of the average value and standard deviation per measurement dataset is shown in Fig.~\ref{fig:supp_average_params}. How these values are used for further calculations is explained in the next section.

\section{Model of coincidence probabilities}
\label{sec:supp_ModelofCoincidence}
In this section we will give details on the model used to calculate the expected coincidence probabilities assuming both completely distinguishable photons and (partially) indistinguishable photons with an indistinguishability $\eta$. We collect coincidences in 19 distinct difference bins (-9 ... 9) which are generated as follows: Once both nodes have indicated their ready-state we perform 10 pulsed excitations. This results in 10 consecutive time windows in which we can detect single photons emitted from the NV centers. For each detected photon we are looking for photon coincidences (i.e. events for which also the other detector clicked). We consider coincidence events where both photons were detected in the same time window (these coincidences are shown in the zero-difference bin) as well as photon detections in different time windows. Depending on the difference in time window number we assign these coincidences to one of the 19 possible bin number differences, numbered from $-9$ to $9$, as shown in Fig.~\ref{fig:Figure4}a of the main text. Due to the finite number of time windows, the probability of detecting a coincidence decreases for higher bin number differences. The probability of detecting a coincidence for any given bin number difference is thus given by $P_{det} = s P_{coinc}$ with the scaling factor $s=10-|bin|$ where $bin$ is the bin number difference. 
The probability of detecting a coincidence is:
\begin{equation}
\label{eq:supp_total_prob}
P_{coinc}=P_{NV NV} + P_{NV DC} + P_{DC DC}
\end{equation}
where $P_{NV NV}$ is the probability of a coincidence between two photons emitted by an NV center, $P_{NV DC}$ is the probability of a coincidence between an NV emitted photon and a dark count (or other noise or background count) and $P_{DC DC}$ is the probability of a coincidence between two dark counts. 

For the non-zero bin number differences these probabilities are given by
\begin{eqnarray}
\label{eq:supp_prob_dist}
P_{NV NV} &= &p_{1A}p_{2B}+p_{1B}p_{2A} + p_{1A}p_{1B} + p_{2A}p_{2B}\\
\label{eq:supp_prob_NVDC}
P_{NV DC} &= &p_{DC_A}(p_{1B}+p_{2B}) + p_{DC_B}(p_{1A} + p_{2A})\\
\label{eq:supp_prob_DCDC}
P_{DC DC} &= &p_{DC_A}p_{DC_B}
\end{eqnarray}
where $p_{1(2)A(B)}$ is the probability of a photon emitted by node 1 (2) to be detected in detector A (B) and $p_{DC_A(B)}$ is the probability of a dark count in detector A (B). All these parameters are measured during the experiment by having excitation rounds from only a single node interleaved with the normal excitation rounds. 

As the single emitter nature of our NV center does not permit two photons from the same NV within one time window (up to a small probability of double excitation which we neglect in this model), in the zero bin number difference $P_{NVNV}$ reduces to the following expression: 
\begin{equation}
\label{eq:supp_NVNV0bin}
P_{NV NV, 0bin} = (p_{1A}p_{2B}+p_{1B}p_{2A})(1-\eta) 
\end{equation}while $P_{NVDC}$ and $P_{DCDC}$ remain the same. Here $\eta$ is the indistinguishability, which represents the reduction of the probability of getting an coincidence from NV contributions. This single parameter in our model takes into account the possible sources of distinguishibility such as spectral/temporal offset and polarization differences.

\section{Calculating the visibility with unbalanced emitters}
We define Visibility using the ratio between coincidence counts in the case that photons are fully distinguishable $C_{dist}$ and coincidence counts we measured, $C_{M}$.
\begin{equation}
V = 1 - \dfrac{C_{M}}{C_{dist}}
\end{equation}
We can directly extract $C_M$ from our experiment, as it corresponds to the measured coincidences in the zero-difference bin. In our model, it is given by filling in Eq.~\ref{eq:supp_NVNV0bin} and \ref{eq:supp_prob_dist} into Eq.~\ref{eq:supp_total_prob}:
\begin{equation}
\label{eq:supp_PM}
P_M = (p_{1A}p_{2B}+p_{1B}p_{2A})(1-\eta) + p_{DC_A}(p_{1B}+p_{2B}) + p_{DC_B}(p_{1A} + p_{2A}) + p_{DC_A}p_{DC_B}
\end{equation}

However, we do not have such direct access to $C_{dist}$. To extract $C_{dist}$ from measured parameters we fit the number of coincidences in the non-zero bins using a linear fit (as shown in Figure 4 of the main text), and use this to extrapolate the value in the zero-bin, which we will call $C_E$. This extrapolation will overestimate the coincidences we would get in the zero bin number difference for perfectly distinguishable photons as the non zero bin number differences also include coincidences of photons emitted from the same NV, a case that does not occur in the zero bin number difference. To determine the necessary correction we compare the probability of a coincidence in the zero bin number difference for $\eta = 0$
\begin{equation}
P_{dist}=(p_{1A}p_{2B}+p_{1B}p_{2A}) + p_{DC_A}(p_{1B}+p_{2B}) + p_{DC_B}(p_{1A} + p_{2A})+ p_{DC_A}p_{DC_B} 
\end{equation}
with $P_{E}$, the probability for an extrapolated zero-bin using the fit of the non-zero bins
\begin{equation}
\label{eq:supp_PE}
P_{E}=(p_{1A}p_{2B}+p_{1B}p_{2A}) + (p_{1A}p_{1B} + p_{2A}p_{2B}) + p_{DC_A}(p_{1B}+p_{2B}) + p_{DC_B}(p_{1A} + p_{2A})+ p_{DC_A}p_{DC_B}
\end{equation}
Resulting in the following correction for the $C_E$, where we use $C_E= P_{E} N_{attempt}$  
\begin{equation}
   C_{dist} = C_E - (p_{1A}p_{1B} + p_{2A}p_{2B})N_{attempt}
\end{equation}

Here we have used the number of experimental attempts $N_{attempt}$ (i.e. the number of excitation pulses sent to the NV center over the course of the entire experiment).
We note that in case the used beamsplitter is strongly imbalanced this would also affect the number of coincidences in the zero bin number difference. In particular, our determined indistinguishability would be scaled by an additional correction factor to $\eta(1-\dfrac{(R-T)^2}{(R^2 + T^2)})$ with the beam splitter reflectivity $R$ and transmitivity $T$.  Using the specified values of our beam splitter  $R=0.496$ and $T=0.504$, we find that this effect is on the order of $\eta(1-10^{-4})$ and we therefore neglect it in our model.

\section{Model of temporal shape of distinguishable coincidences}
\label{sec:supp_ModelofTemporal}
The analysis of the measured coincidences with respect to their arrival time difference holds a vast amount of information of the (relative) emitter properties. In our case, it is difficult to extract detailed information about the NV centres from the measured coincidences in the zero-difference bin as the vast majority of these coincidences involves background counts. We can, however, model the different temporal shape of the three contributions as mentioned in \ref{eq:supp_total_prob} in the non-zero difference bins.
Analogous to Kambs~\cite{Kambs_limitationsTPQI_2018} we start by writing down the photon wave-packets of the individual events. The single photon wave function an NV-centre from spontaneous emission (ignoring any spectral and phase information) is given by

\begin{equation}
    \phi_{NV}(t) = \frac{1}{\sqrt{\tau(e^{\frac{-T_{start}}{\tau}}-e^{\frac{-T_{end}}{\tau}}})}e^{-\frac{t}{2\tau}}H(t-T_{start})(1-H(t-T_{end}))
\end{equation}

where $\tau =$\SI{12.5}{\nano\second} the lifetime of the excited state. We can define $W(t) \equiv H(t-T_{start})(1-H(t-T_{end})$ to be the selected window with start(end) time $ T_{start}$($T_{end}$) both larger than $0$. $H(t)$ is the Heaviside step function. Here we have chosen $t = 0$ to be the moment the infinitely short optical excitation would have arrived in the detector as the start of the exponentially decaying wave-packet. We now assume that both NV centres have the same excited state lifetime, and the beamsplitter ratio to be perfect. The joint detection probability is given by:

\begin{equation}
    P_{joint}(t, \Delta t) = \frac{1}{4}\lvert\phi_i(t + \Delta t)\phi_j(t) - \phi_j(t + \Delta t)\phi_i(t)\rvert^2
\end{equation}

A background photon at the beamsplitter is modeled as a process that has a uniform probability density in time:
\begin{equation}
    P_{DC}(t) = \frac{1}{(T_{end}-T_{start})}W(t)
\end{equation}
 Here we assume that the contribution of background noise is constant in time, which is a good approximation for the darkcounts of the detector and SPDC noise coming from the QFC process. For the excitation pulse leakage this approximation does not hold, but we assume this error to be small if we limit the amount of pulse light in our analysis window. For completely distinguishable events the interference term in the joint detection probability drops out, and it simplifies to:

\begin{equation}
    P_{joint}^{dist}(t, \Delta t) = \frac{1}{4}(P_i(t + \Delta t)P_j(t) + P_j(t + \Delta t)P_i(t))
\end{equation}

with $P_i(t) = \lvert\phi_i(t)\rvert^2$. To obtain the second order cross-correlation function we integrate over the detection times of the single photons. The NV-NV contribution is therefore given by:

\begin{eqnarray}
 \mathcal{G}_{NVNV}^{(2)}(\Delta t) = \int_{-\infty}^{\infty}\frac{1}{4}\ 2 P_{NV}(t') P_{NV}(t'+\Delta t) dt' =\\
 \frac{1}{2\tau^2(e^{\frac{-T_{start}}{\tau}}-e^{\frac{-T_{end}}{\tau}})^2}\int_{-\infty}^{\infty}e^{-\frac{t'}{\tau}}e^{-\frac{t'+\Delta t}{\tau}}W(t')W(t'+\Delta t) dt'
\end{eqnarray}
We can solve this integral by breaking it up for positive and negative $\Delta t$, and absorbing the Heavisides accordingly in the integration limits. For $\Delta t > 0$ we get:

\begin{equation}
 \frac{1}{2\tau^2(e^{\frac{-T_{start}}{\tau}}-e^{\frac{-T_{end}}{\tau}})^2}\int_{T_{start}}^{T_{end} - \Delta t}(e^{-\frac{t'}{\tau}}e^{-\frac{t' + \Delta t}{\tau}}) \,dt' = \frac{1}{4\tau(e^{\frac{-T_{start}}{\tau}}-e^{\frac{-T_{end}}{\tau}})^2}(-e^\frac{-2T_{end} - \Delta t}{\tau} + e^\frac{-2T_{start} + \Delta t}{\tau})
\end{equation}

For $\Delta t < 0$ we get:

\begin{equation}
  \frac{1}{2\tau^2(e^{\frac{-T_{start}}{\tau}}-e^{\frac{-T_{end}}{\tau}})^2}\int_{T_{start} - \Delta t}^{T_{end}}(e^{-\frac{t'}{2\tau}}e^{-\frac{t' + \Delta t}{2\tau}}) \,dt' = \frac{1}{4\tau(e^{\frac{-T_{start}}{\tau}}-e^{\frac{-T_{end}}{\tau}})^2}(-e^\frac{-2T_{end} + \Delta t}{\tau} + e^\frac{-2T_{start} - \Delta t}{\tau})
\end{equation}

which, after combining both, arrive at the expression for $\Delta t \in [-(T_{end}-T_{start}), T_{end}-T_{start}]$:

\begin{equation}
\mathcal{G}_{NVNV}(\Delta t) = \frac{1}{4\tau(e^{\frac{-T_{start}}{\tau}}-e^{\frac{-T_{end}}{\tau}})^2}(e^\frac{2T_{start} + |\Delta t|}{\tau} - e^\frac{2T_{end} - |\Delta t|}{\tau})
\end{equation}

As a sanity check, we can integrate $\mathcal{G}_{NVNV}(\Delta t)$ over the interval $[-(T_{end}-T_{start}), T_{end}-T_{start}]$, to get the overall probability to get a coincidence in our window:

\begin{equation}
    \int_{-(T_{end}-T_{start})}^{T_{end}-T_{start}}\mathcal{G}_{NVNV}(\Delta t)d\Delta t
\end{equation}
which, by separately integrating the for negative and positive $\Delta t$ results in
\begin{equation}
\frac{1}{4\tau(e^{\frac{-T_{start}}{\tau}}-e^{\frac{-T_{end}}{\tau}})^2}\tau\left((e^{\frac{-T_{start}}{\tau}}-e^{\frac{-T_{end}}{\tau}})^2 +( e^{\frac{-T_{start}}{\tau}}-e^{\frac{-T_{end}}{\tau}})^2\right) = \frac{1}{2}
\end{equation}
This is what we would expect for two completely distinguishable photons.

Following the same approach, we can calculate the integral for the NV-DC as 

\begin{equation}
    \mathcal{G}_{NVDC}^{(2)}(\Delta t) =\int_{T_{start}}^{T_{end}}\frac{1}{4}(P_{NV}(t' + \Delta t)P_{DC}(t') + P_{DC}(t' + \Delta t)P_{NV}(t')) \,dt' =
\end{equation}

\begin{equation}
\frac{e^\frac{-T_{start}}{\tau} - e^{-\frac{T_{end} - |\Delta t|}{\tau}} + e^{-\frac{T_{start} + |\Delta t|}{\tau}} - e^\frac{-T_{end}}{\tau} }{4\tau(T_{end}-T_{start})(e^{\frac{-T_{start}}{\tau}}-e^{\frac{-T_{end}}{\tau}})}
\end{equation}

For the DC-DC contribution we get

\begin{equation}
    \mathcal{G}_{DCDC}^{(2)}(\Delta t) =\int_{T_{start}}^{T_{end}}\frac{1}{4}(P_{DC}(t' + \Delta t)P_{DC}(t') + P_{DC}(t' + \Delta t)P_{DC}(t')) \,dt' =  \frac{(T_{end}-T_{start}) - |\Delta t|}{2(T_{end}-T_{start})^2}
\end{equation}
that both integrate to $\frac{1}{2}$ over the coincidence window. 

Figure~\ref{fig:supp_temp_shapes} shows the shapes of these cross-correlation functions for the different contributions of coincidences in the experiment. The difference between NV-DC and DC-DC contributions is minimal, but the NV-NV shape is clearly distinguishable from the noise sources. The plot in Fig.~\ref{fig:Figure4}b in the main text is a weighted sum of the three shapes. The weights of for the different contributions are calculated using Eq. ~\ref{eq:supp_prob_dist}-~\ref{eq:supp_prob_DCDC} and the data shown in Fig.~\ref{fig:supp_average_params}. For the numerical values of input parameters used see Tab. ~\ref{tab:supp_overview_numerical_values} For the indistinguishable case, we use the same temporal shape as the distinguishable, and scale the amplitude with $1-\eta$. This is a valid approximation where polarization difference are the dominant contribution with respect to any temporal or spectral effects.

\section{Monte-Carlo based derivation of Indistinguishability using Bayesian Inference}
\label{sec:supp_bayesian_inference}
\begin{figure}[ht]
    \centering
	\includegraphics[scale = 0.3]{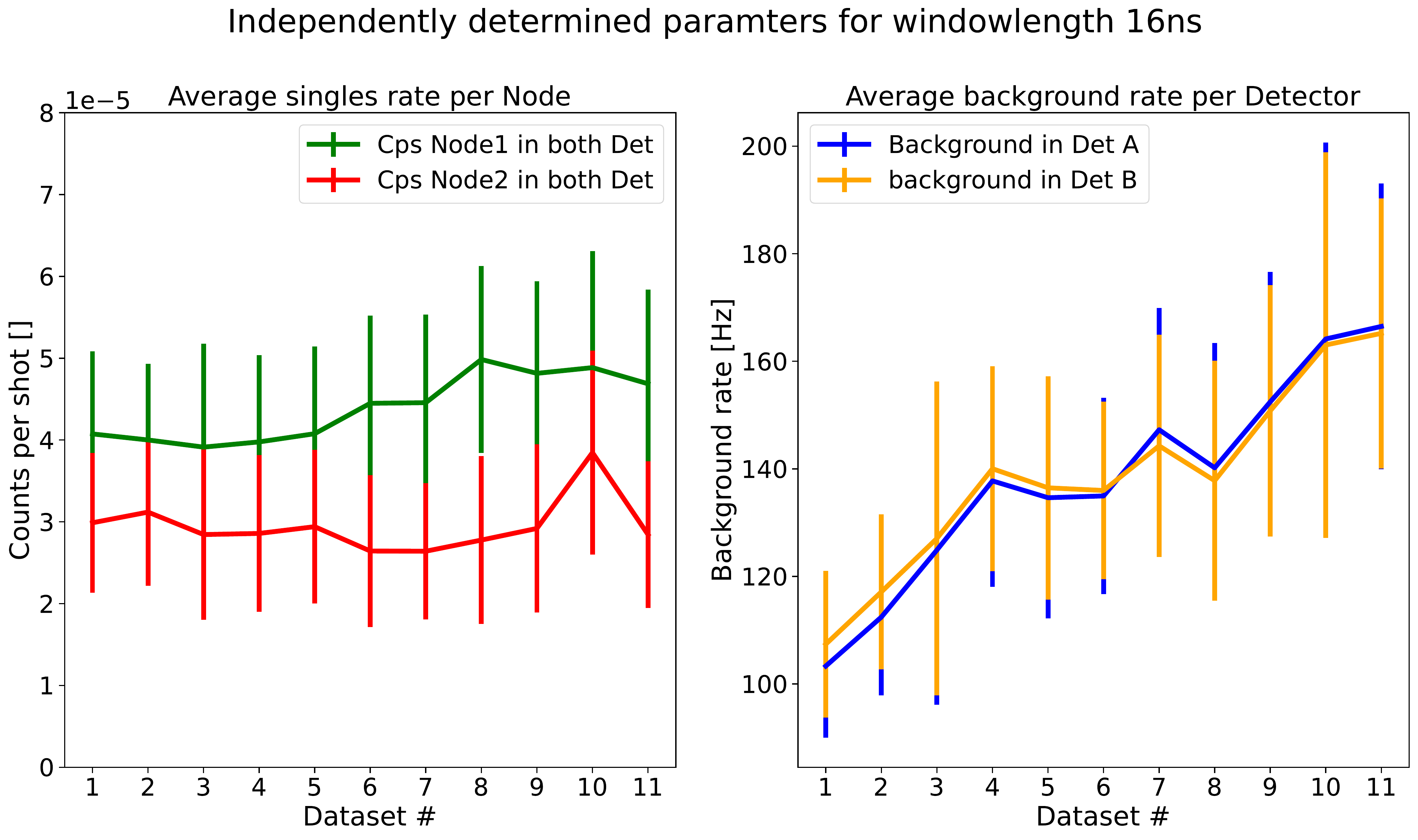}
	\caption{\label{fig:supp_average_params}Average of the independently measured parameters for one of the analyzed integration windows. Individual data sets are about $24$ hours of measurement. The error-bar on the values is the standard error of the mean. For each datapoint,the parameter is measured during at approximately 2 minute intervals. The signal rate is calculated in the fluorescence window after the excitation as shown in figure \ref{fig:Figure3}. The average background rate is calculated using data in between the pulses, far away from the optical excitation and fluorescence. These parameters can be calculated for the different windowlengths using the same dataset. The rise of the background counts during the experiment is currently not understood.}
\end{figure}

As a next step we use our model to derive an expression for the indistinguishablity $\eta$. Due to the stochastic nature of the coincidences observed in our measurement, the calculation of the indistinguishability using standard error propagation like the variance formula can lead to non-physical results for the indistinguishability being in its confidence interval. We prevent this by using Bayesian Inference to find a more accurate confidence interval. In words, we want to know the most likely indistinguishability $\eta_{opt}$, given the set of measured coincidences $C_{M}$ and $C_{E}$. Equations~\ref{eq:supp_PM} and~\ref{eq:supp_PE} and the number of attempts $N_{attempts}$ give us a direct way of calculating $C_{M}$ and $C_{E}$, based on $p_{1(2)A(B)}$, $p_{DC_{A(B)}}$ (denoted $\bar{\theta}$ in short) and $\eta$ as described in~\ref{sec:supp_ModelofCoincidence}.
The strategy is then to simulate many realizations of our experiment with $\eta_i$ the only free parameter, via a Monte-Carlo simulation. We then use the outcomes of these simulations to calculate the likelihood of a certain $\eta$, given our measured $C_{M}$ and $C_{E}$ via Bayes rule:

\begin{equation}
\label{eq:supp_BayesRule}
P(\eta, \bar{\theta} | \bar{M} ) = \frac{P(\bar{M}|\eta, \bar{\theta})P(\eta, \bar{\theta})}{P(\bar{M})} = \frac{P(\bar{M}|\bar{\theta},\eta)P(\eta, \bar{\theta})}{\sum_i P(\bar{M}|\bar{\theta}, \eta_i)P(\eta_i, \bar{\theta}) }
\end{equation}

where $P(a|b)$ is the probability of $a$ given $b$, $\bar{M}$ is the observed measurement outcomes $C_{M}$ and $C_{E}$, and $\bar{\theta}$ the vector containing the parameters that fully describe our experiment. $P(\eta, \bar{\theta})$ is the probability of the set of parameters \textit{before} our measurement, the so called prior. The only unknown parameter is $\eta$, for which we take a uniform distribution on the interval $[0,1]$, to reflect the assumption of no prior knowledge. All the other parameters $\theta_{j}$ in $\bar{\theta}$ are assumed to be normally distributed, with the mean and variance determined by the independent samples during the measurement, see previous section~\ref{sec:supp_DataCollection} and~\ref{fig:supp_average_params}. 

\begin{algorithm}[H]
    
    \caption{\label{alg:supp_BayesianEstimation}Monte-Carlo simulation routine to find likelyhood of $\eta_{i}$, based on the inputs $\bar{\theta}$ and $\bar{M}$}
	\begin{algorithmic}[1]
  		\For {$\eta_i$ in $[0 \ldots 1]$}
			\For {$\theta_j$ in $\bar{\theta}$}
				\State Draw $N$ times input parameters $\mathbf{\theta_{n}}$ from $\mathcal{N}(\mu_j,\,\sigma_j^{2})$
				\State Calculate N values for $P_E^n$ and $P_M^n$ according to eq. S8 and S10
			\EndFor
			\For {$n$ in $[0, 1, \ldots, N]$}
			    \State Draw $K$ realizations of the measurement outcomes  
			    \State $C_E^{Sim} = Poisson(P_E^n * N_{attempts})$ and $ C_N^{Sim} = Poisson(P_M^n * N_{attempts})$
			    \State Store $C_E^{Sim}$ and $C_M^{Sim}$ in array
			\EndFor
             \State Calculate fraction $\frac{\sum{[(C_E^{Sim} = C_E) \land (C_M^{Sim} = C_M)}]}{N*K}$, which is equal to $P(\bar{M}|\eta, \bar{\theta})$
        \State Calculate likelyhood of $\eta_i$ using eq. S23 and store the value.
	    \EndFor
	\end{algorithmic}
\end{algorithm}

\begin{figure}[ht]
    \centering
	\includegraphics[scale = 0.4]{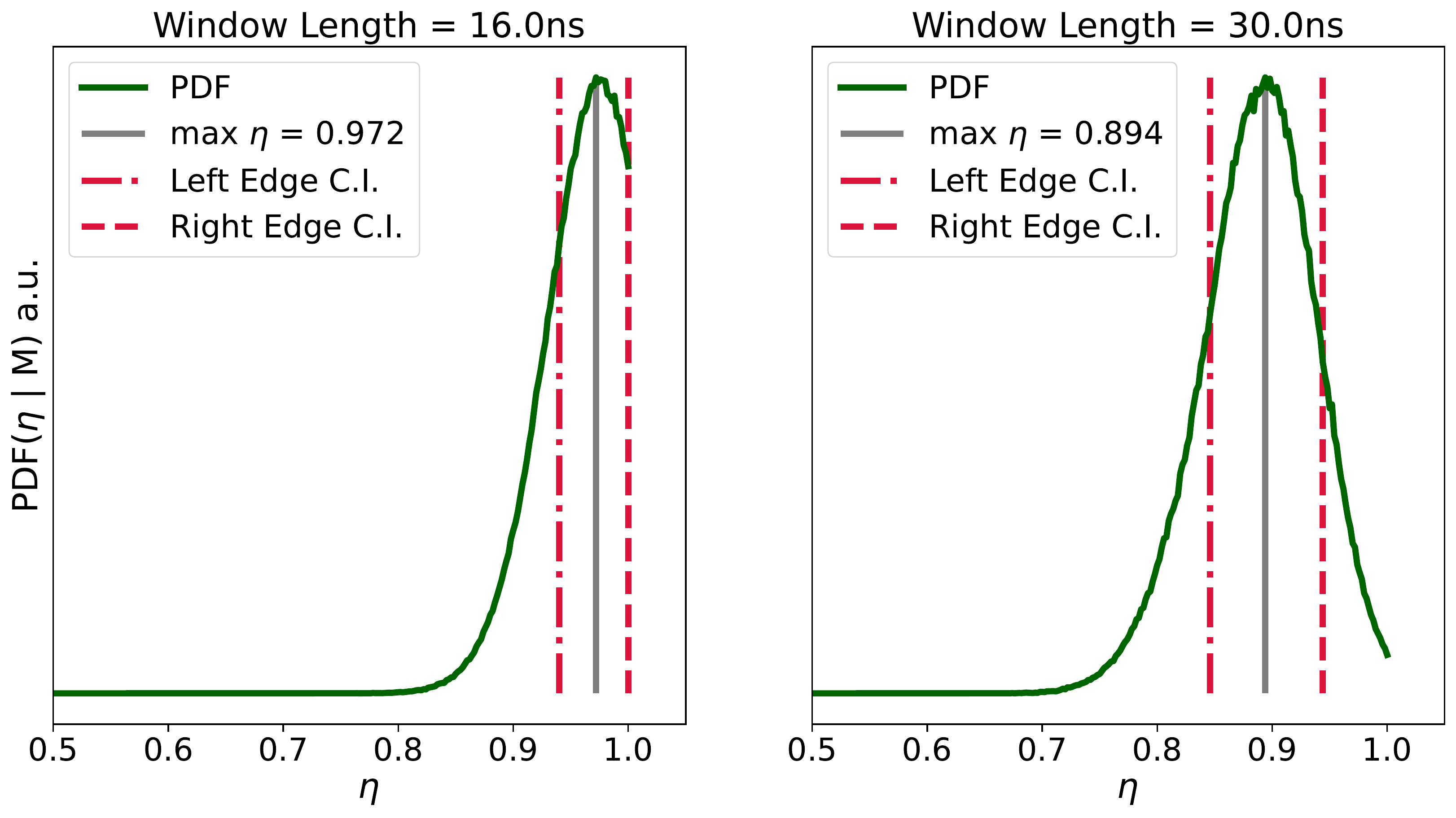}
	\caption{\label{fig:supp_pdf_indist}Probability density functions generated by the Monte-Carlo simulations, used for the calculation of the indistinguishability shown in the main text and figure\ref{fig:Figure4}. A clear optimum is seen for both the distributions. The asymmetric confidence intervals are the result of the distribution being close to $1$, as shown by the distribution for the left figure.}
\end{figure}

The algorithm to calculate the posterior distribution of $\eta$ for a given set of $\bar{\theta}$ and $\bar{M}$ is given in algorithm~\ref{alg:supp_BayesianEstimation}. It captures both the uncertainty in the input parameters $\bar{\theta}$, as well as the statistical fluctuations introduced by the stochastic nature of measuring a rate of coincidences. In our case, the uniform chosen prior means that the normalized likelyhood function produced by the algorithm is directly the probability density function (pdf) for the left hand side of Eq.~\ref{eq:supp_BayesRule}. The pdf's for the calculated indistinguishability of the data shown in Fig.~\ref{fig:Figure4}a and b are shown in Fig.~\ref{fig:supp_pdf_indist}. Here we can clearly see the asymmetry of the calculated distributions, and their cut-off at the maximum of 1. To report this likelyhood as a single value with 'errorbars', we chose the maximum of the likelyhood to be our datapoint, and calculate a symmetric $68\%$ confidence interval around the most likely $\eta$. For the datapoints where the most likely indistinguishibility is closest to $1$, a symmetric confidence interval, with $34\%$ of the probability on either side can not be taken, and asymmetric intervals are used.

\section{SNSPD blinding induced background}
\label{sec:supp_OpusBlinding}
The current lay-out of the in-fibre optics guides the reflected part of the reference light through the same port of the circulator as the single-photons. Therefore the power is aimed directly at the nanowires of the SNSPDs, blinding the dectectors. Additionally, because of the beamsplitter in front of the detectors, the reference light reflected from both UNFs interferes, making the output power oscillate in time. To prevent latching, the manufacturer placed anti-latching shunt resistors in the circuits to prevent the loss of photon counting over timescales of $>$milliseconds.

\begin{figure}[ht]
    \centering
	\includegraphics[scale = 0.6]{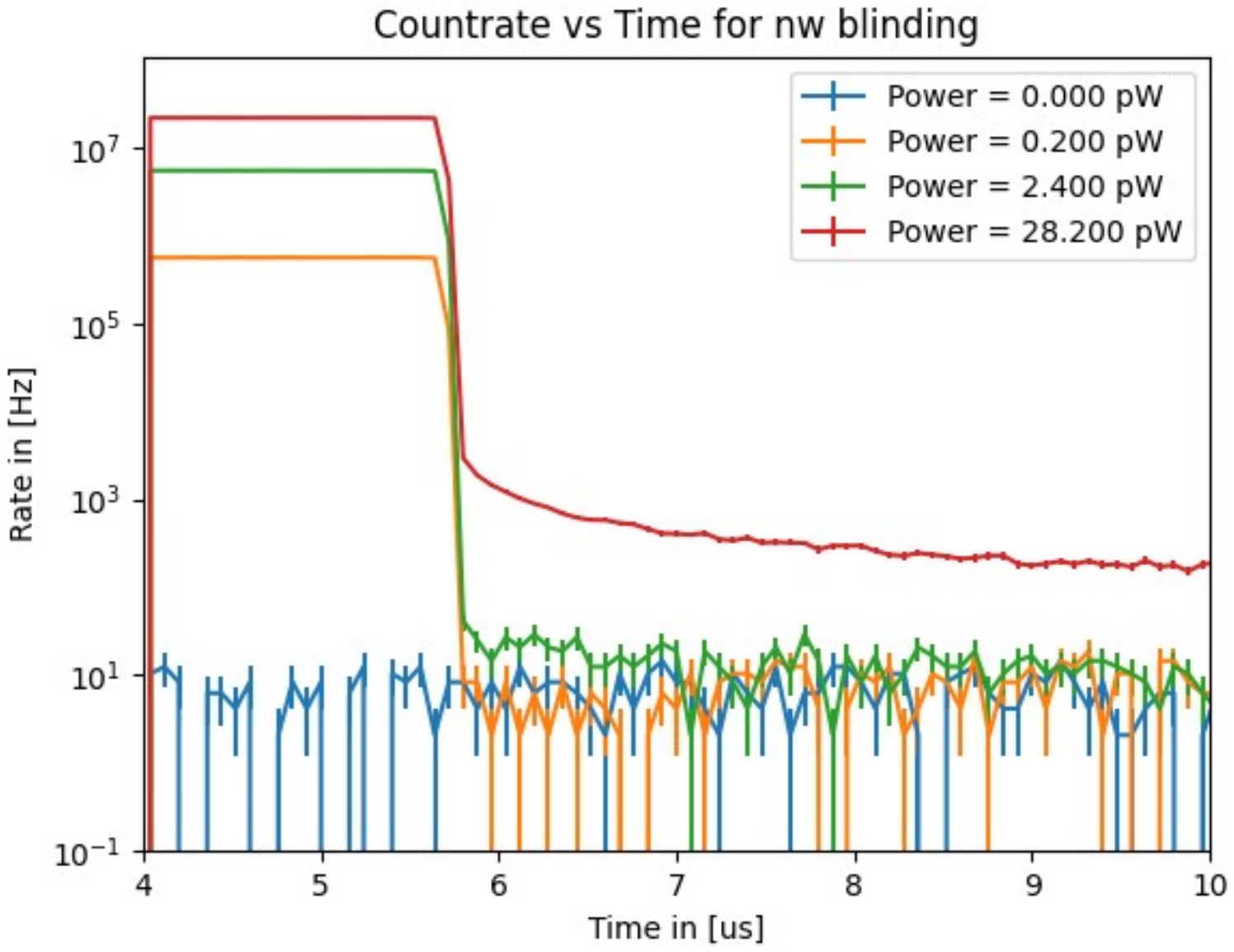}
	\caption{\label{fig:supp_opus_blinding}Rate of measured events after a bright blinding pulse in the superconducting nanowire single photon detectors. After the bright pulse we measure an elevated background level as compared to no blinding pulse applied. This effect persists for tens of microseconds, forcing us to delay the single photon generation by more than \SI{45}{\micro\second} with respect to the frequency stabilization measurement. The optimal powers used during the experiment is not shown, and was optimized in-situ.}
\end{figure}

However, during the early investigation of the time multiplexing of frequency stabilization with single photon generation, we noticed a blinding-power dependent elevated darkcount rate, persisting long ($>$\SI{10}{\micro\second} after the end of the blinding pulse. By scanning the incident power on one of the nanowires in a controlled manner (Fig.~\ref{fig:supp_opus_blinding}, we can see a clear power dependence of this effect. We reduce the additional background counts to a constant manageable level of \SI{30}{\hertz} by moving our single photon generation by $\approx$\SI{45}{\micro\second}, and optimizing the reference light power in-situ. Further improvement to the in-fibre optics that reduce the backpropagated power through the FBG and active attenuation shielding the detectors can be employed in the future to remove this background contribution.

\section{UNF stability}
\label{sec:supp_UNF_stability}
The UNFs are enclosed in an well-isolated box, with a heating pad and thermometer, temperature controlled by a Team Wavelength TC5. This controller provides $\sim$\SI{0.1}{\milli\kelvin} control of the temperature setpoint. Due to non-homogeneous distribution of temperature inside the box and fluctuating temperature gradients coming from outside the box, with a fixed temperature setpoint we observe significant drift of the UNF frequency if not actively stabilized to a target frequency.\\

The UNFs are frequency stabilized by measuring the power incident on the detector (Thorlabs PDB482C-AC) that measures the frequency lock beat, extracted as a voltage from a monitoring port of the detector. We stabilize the temperature controller setpoint of the UNF to the relative half-of-maximum transmission point. We calibrate the maximally measured power through each individual UNF by temperature-sweeping them, effectively changing the position of the filters in frequency space. In this calibration, there is no correction for the noise incident on or coming from the detector, . After calibration, we feedback the measured transmission power back to the temperature setpoint of the temperature controller of the UNF with a software implemented PI-control loop.

\subsection{Transmission stability}
 Due to inherent inaccuracies in the active stabilization described above, a spread of transmission powers of the UNFs with respect to their setpoint at the relative half-of-maximum transmission point is observed. This occurs due to the limitation in accuracy of the actual temperature control at the filter and other drifts of fixed in-fiber components in the path that passes through the UNF. Such a drift in frequency of the UNF maximum transmission point results in a reduction of NV photon transmission probability, however the NV conversion target frequency is unaffected. To determine the exact UNF frequency shift with respect to the setpoint, we start with the Cauchy-distribution fit in Fig.~\ref{fig:Figure2}d of the transmission power $T$ of the reference laser through the filter with respect to frequency $f$:
\begin{equation}
    T\left(f\right) = \frac{I}{1 + \left(\frac{\left(f-f_0\right)}{\gamma}\right)^2} + B,
    \label{eq:cauchy_fbg}
\end{equation}
where $f_0$ is the location of the filter, $2\gamma$ the filter FWHM and $I$ and $B$ a relative scaling factor and offset to background, respectively. We can relate the offset from the half-of-maximum transmission point to a frequency shift on this characteristic shape of the filter by rewriting the above function to be a function of transmission
\begin{equation}
    f = f_0 - \gamma\sqrt{-1+\frac{I}{\left(T\left(I+B\right)-B\right)}},
    \label{eq:inverted_cauchy}
\end{equation}
 obtaining $f$ by using the filter-specific fit parameters obtained from Eq.~\ref{eq:cauchy_fbg}. The root-minus solution is chosen according to the boundary condition that appropriately reflects the direction of change in frequency for a change in transmission power. With Eq.~\ref{eq:inverted_cauchy} we can convert a representative stabilized \SI{70}{\hour} set of second-interval transmission power data to a relative shift in frequency of the filter with respect to the the half-of-maximum transmission point, in the assumption that all transmission power drift is due to the shift in UNF frequency. The transmission power is not corrected for any other drifts of in-fiber components between the power-stabilized reference laser emission and detection at the beat detector. The residual frequency instability for both filters is shown in Fig.~\ref{fig:supp_UNF_transmission}.


\subsection{NV photon transmission}
From this frequency instability we can calculate the change in average transmission probability of converted photons emitted from the NV center through the UNFs. The emitted NV photons with natural linewidth of $2\gamma_{\mathrm{NV}}\approx$\SI{12.7}{\mega\hertz}~\cite{Goldman_NVlifetime_2015} have a frequency distribution as the Cauchy distribution
\begin{equation}
    P\left(f\right) = \frac{1}{\pi\gamma_{\mathrm{NV}} \left(1 + \left(\frac{\left(f-f_{NV}\right)}{\gamma_{\mathrm{NV}}}\right)^2\right)},
\end{equation}
where $f_{NV}$ is the frequency of our target wavelength (see Fig.~\ref{fig:Figure2}). We can treat the UNFs as frequency dependent transmission devices, i.e. the Cauchy fit parameters used for Fig.~\ref{fig:Figure2}D can be used to construct the \textit{probability} of transmission $T_{\mathrm{p}}\left(f\right)$ of an incident photon by setting $I=1, B=0$ in Eq.~\ref{eq:cauchy_fbg}, resulting in $T\left(f=f_0\right) = 1$. The probability of the NV photon transmitting through the UNF filter is then
\begin{equation}
    P_{\mathrm{t}} = \int_{-\infty}^{\infty}P\left(f\right)T_{\mathrm{p}}\left(f\right) \,df.
\end{equation}
Using the above equation we can calculate the relative change in transmission probability from each UNF's fitted parameters and all frequencies $f_0\pm$\SI{1}{\mega\hertz}, the interval where almost all of the occurrences of the measured filter shift are located. Using numerical methods to approximate the above integral we find that the total transmission of the NV photon is reduced to $\approx75\%$(UNF1) and $\approx77\%$ (UNF2) with the filters at the half-of-maximum transmission setpoint, excluding any losses incurred in the physical implementation. For the UNF stability data shown in Fig.~\ref{fig:supp_UNF_transmission} we obtain a transmission change of at most $\approx 1.5\%$(UNF1) and $\approx0.4\%$ (UNF2) for $f_0\pm$\SI{1}{\mega\hertz} that would solely be ascribable to the frequency shift of the UNFs.

\begin{figure}[ht]
    \centering
	\includegraphics[scale = 0.8]{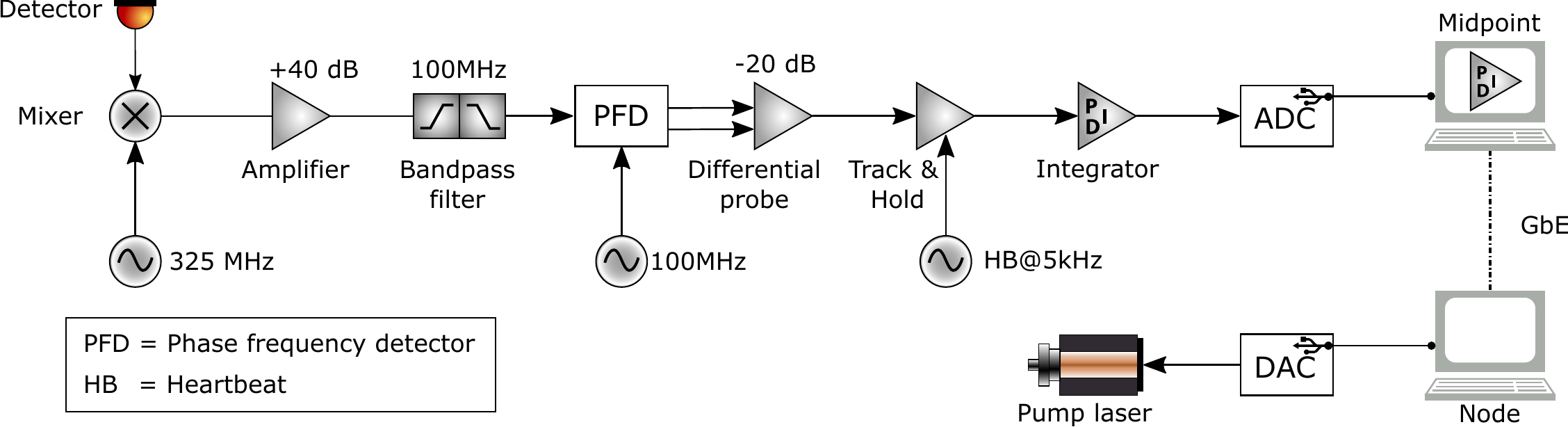}
	\caption{\label{fig:supp_freq_lock} Schematic of frequency lock hardware for one arm of the two-node setup. The optical beat is down-mixed with \SI{325}{\mega\hertz} and then further processed using standard electronics components. The feedback loop is closed using a pair of ADC/DAC over a Gigabit Ethernet connection, resulting in a feedback rate of $\sim \SI{500}{\hertz}$.}  
\end{figure}

\begin{figure}[ht]
    \centering
	\includegraphics[scale = 0.8]{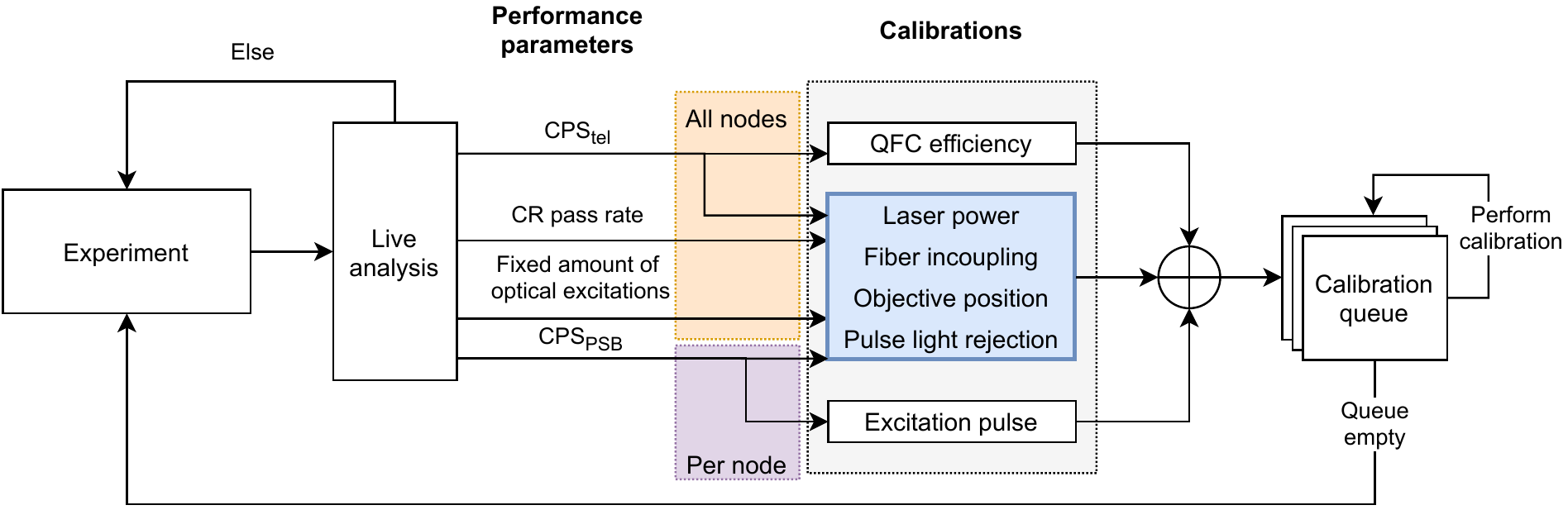}
	\caption{\label{fig:supp_auto_calibrations} Schematic visualization of the automated calibration framework. At fixed intervals of NV excitation attempts, we perform live an analysis of several variables that triggers specific calibrations per node or on all nodes. The order in which the possible calibrations are performed in the queue is the same as the order in which they are shown in the schematic, reading from top to bottom.}
\end{figure}

\begin{figure}[ht]
    \centering
	\includegraphics[scale = 0.8]{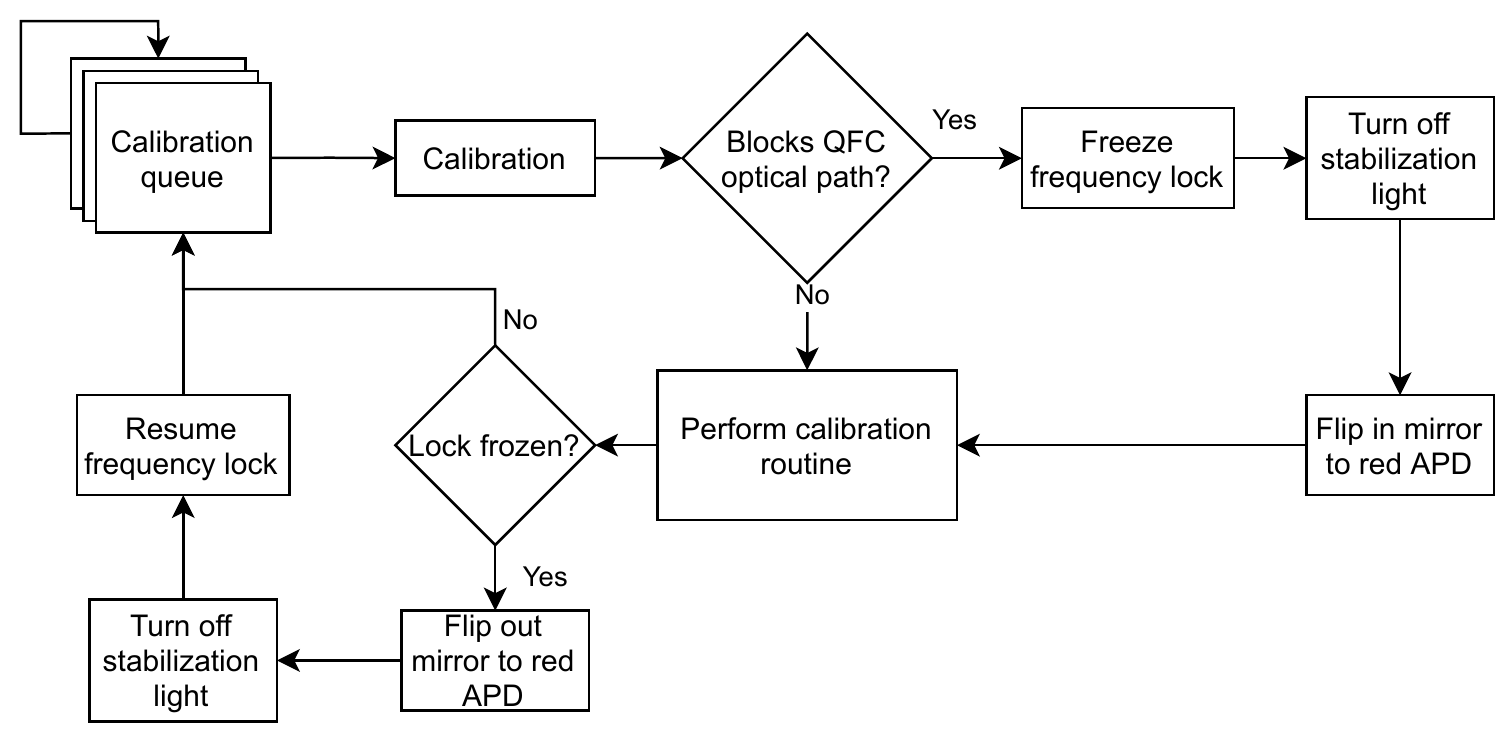}
	\caption{\label{fig:supp_freqlock_calibrations} Schematic showing the decision logic deciding if the frequency lock should be frozen before performing a calibration routine. To ensure proper handoff of setup-equipment control, the stabilization light is always turned off before any calibration is started whilst freezing the feedback, if this calibration will block the optical path going through the QFC. Resuming of the lock is an exact reversal of this process, where the stabilization light has to be turned on before the lock's feedback loop is resumed, otherwise no locking beat will be generated. }
\end{figure}

\begin{figure}[ht]
    \centering
	\includegraphics[scale = 0.5]{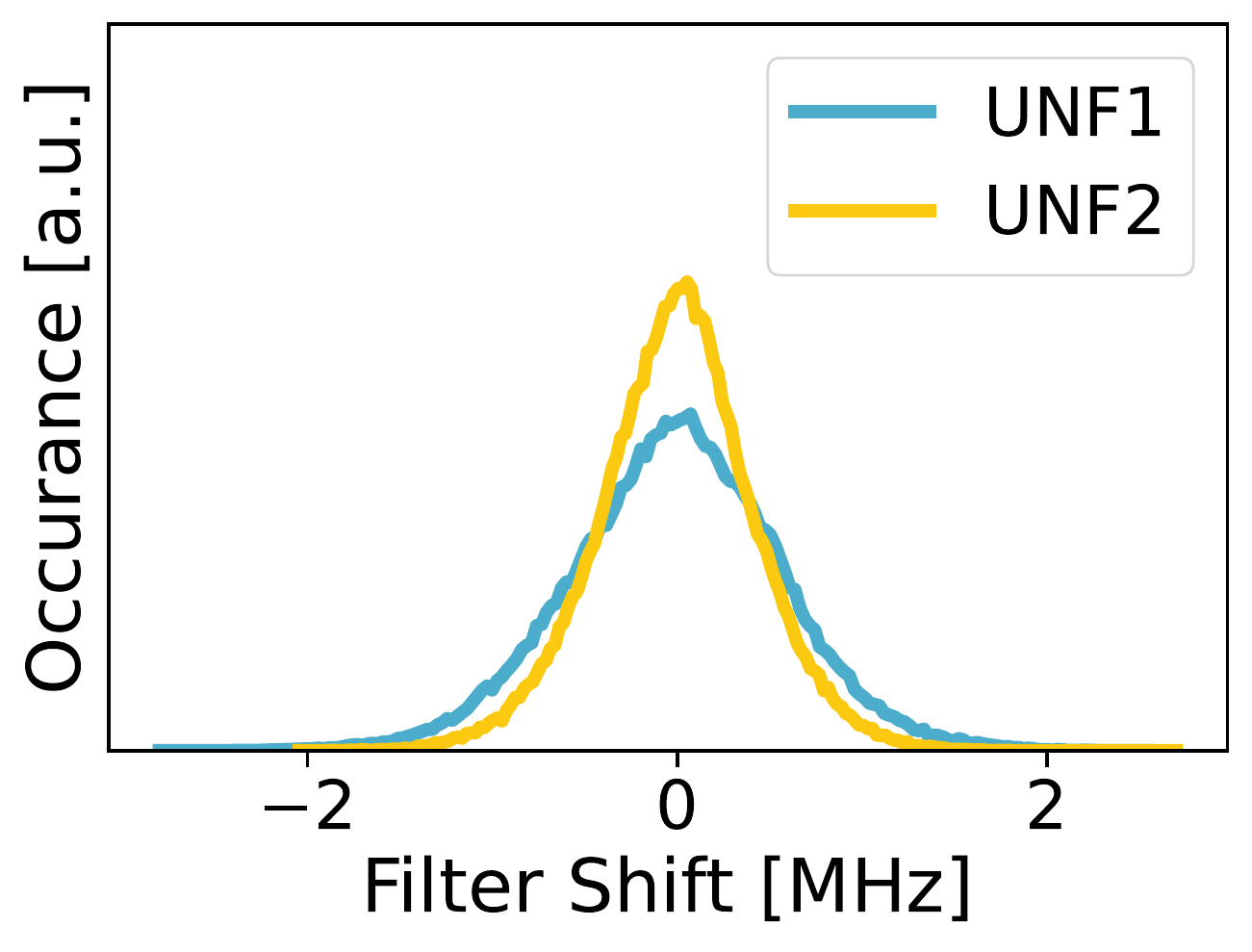}
	\caption{\label{fig:supp_UNF_transmission} Histogram of the calculated filter shift of both UNFs with respect to their half-of-maximum transmission setpoint over a representative frequency-stabilized \SI{70}{\hour} set of second-interval transmission power measurements, in the assumption that all transmission power drift is due to the shift in UNF frequency. From the NV photon and filter overlap we calculate to have at most a change of $\approx1.5\%$ in transmission power for a change in filter frequency of $\pm$\SI{1}{\mega\hertz}, covering approximately 92\% and 97\% of the spread in filter shift, for UNF1 and UNF2, respectively.}
\end{figure}

\begin{figure}[ht]
    \centering
	\includegraphics[scale = 0.5]{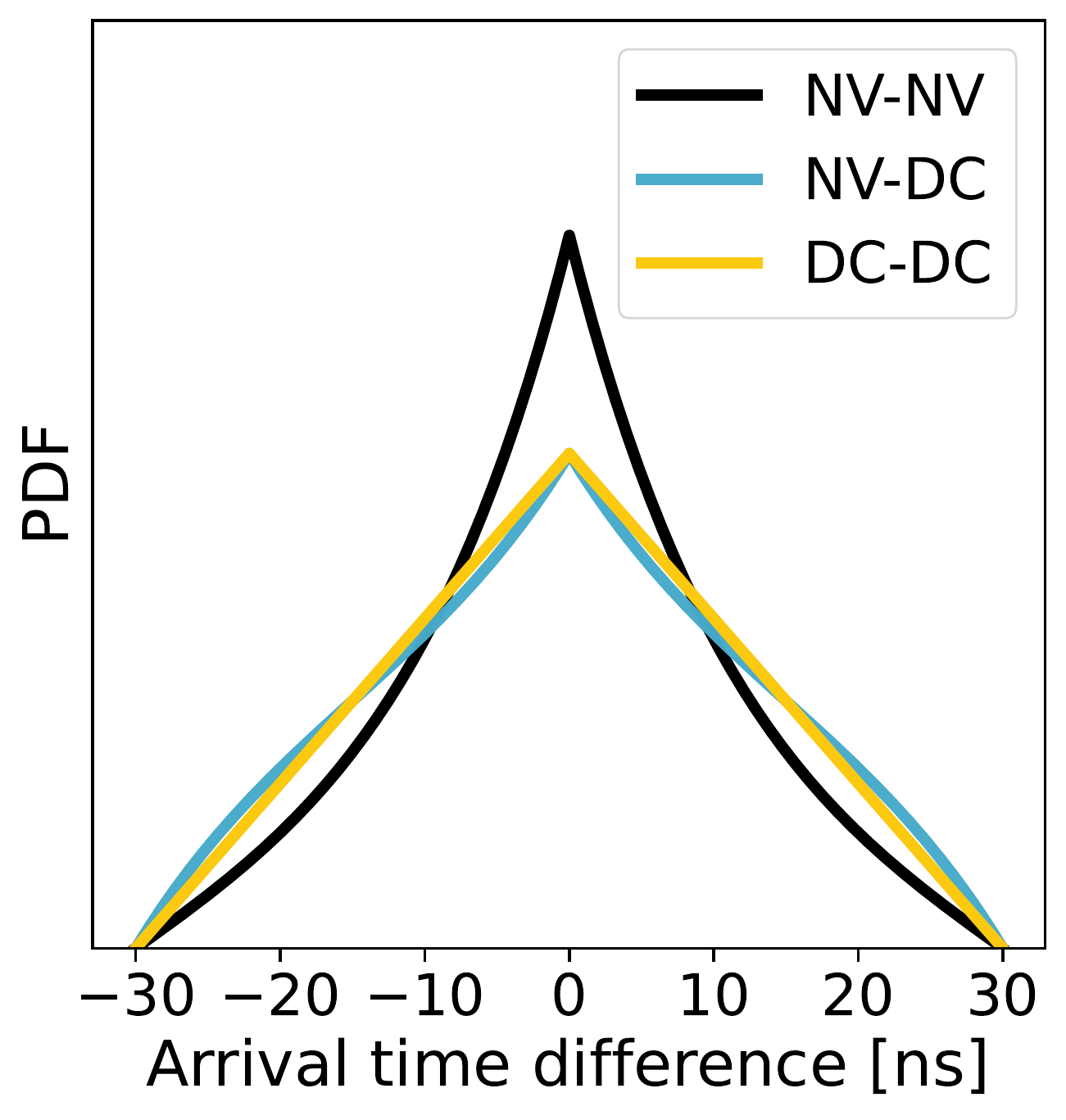}
	\caption{\label{fig:supp_temp_shapes} Temporal shape of the three contributions of our model explained in section \ref{sec:supp_ModelofCoincidence} and \ref{sec:supp_ModelofTemporal}. }
\end{figure}

\begin{figure}[ht]
    \centering
	\includegraphics[scale = 0.5]{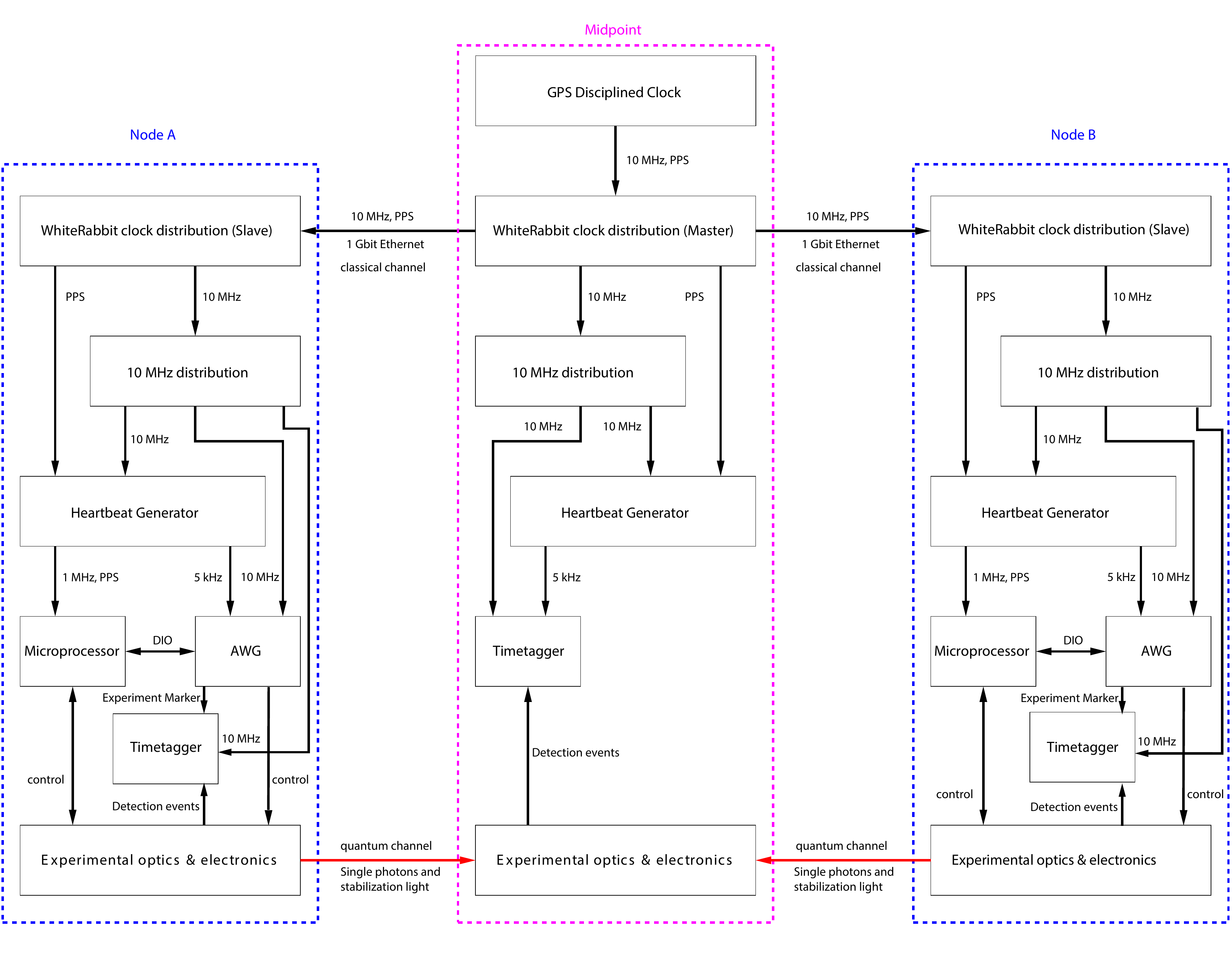}
	\caption{\label{fig:supp_overview_adapted}Schematic overview of hardware to synchronize events across the experiment. All connections that distribute the critical timing signals are propagated over telecom fibres on a dedicated fibre. The data that is generated by the timetaggers is sent over the network to a single PC that processes the data. Both the raw and processed data are stored to disk.}
\end{figure}
\begin{table}[hb]
\centering
\caption{\label{tab:supp_equipment}Experimental equipment used in the two nodes. }
\begin{tabular}{|p{6cm}|p{10cm}|}
\hline
Equipment name & Part name/number   \\ \hline\hline
Cryostat & Montana Instruments Cryostation S50 \\ \hline
Positioners & Microscope objective on 3x PI E-873  \\ \hline
Micro-controller & J\" ager ADwin-Pro II T12 \\ \hline
Arbitrary Waveform Generator $AWG$ & Z\"urich Instruments HDAWG-8\\ \hline
Excitation laser &  Toptica DL pro \SI{637}{nm}  \\ \hline
Spin reset laser & Toptica DL pro \SI{637}{nm} \\ \hline
Charge reset laser &  Cobolt 06-01 \SI{515}{nm}  \\ \hline
Reference laser &  NKT Photonics 1550 nm Koheras ADJUSTIK E15 PM FM \\ \hline
EOM & Jenaoptik AM635b  \\ \hline
AOM & Gooch and Housego Fibre Q 637nm \\ \hline
AOM Driver & Time Base DIM-3000 \\ \hline 
APD &  Laser Component  COUNT-10C Photon Countng Module \\ \hline
SNSPDs & Quantum Opus 00-NPD-1588-HDE  \\ \hline
 \hline
\end{tabular}
\end{table}

\begin{table}[hb]
\centering
\caption{\label{tab:supp_equipment_qfc}QFC equipment used in the two nodes.}
\begin{tabular}{|p{6cm}|p{10cm}|}
\hline
Equipment name & Part name/number   \\ \hline\hline
Pump laser & NKT Photonics 1064 nm Koheras ADJUSTIK Y10 PM FM   \\ \hline
Pump laser amplifier & NKT Photonics Koheras BOOSTIK HP   \\ \hline
Non-linear Crystal & Custom ppLN crystal NTT \\ 
 \hline
Remote Piezo mirrors & Newport AG-M100n \\
\hline
\end{tabular}
\end{table}

\begin{table}[hb]
\centering
\caption{\label{tab:supp_equipment_timing} Timing Hardware }
\begin{tabular}{|p{8cm}|p{8cm}|}
\hline
Equipment name & Part name/number \\ \hline\hline
GPS-disciplined clock & Stanford Research Systems FS752    \\ \hline
Frequency Distribution, between nodes (10MHz, PPS) &  OPNT WRS-3/18 White Rabbit Switch \\ \hline
Frequency Distribution, local (10MHz) & Pulse Research Lab PRL-4110  \\ \hline
Heartbeat Generator & Tektronix AFG 31022  \\ \hline
Micro-controller & J\" ager ADwin-Pro II T12 \\ \hline
Arbitrary Waveform Generator (AWG) & Z\"urich Instruments HDAWG-8\\ \hline
Time Tagger &  PicoQuant MultiHarp 150 4N \\ \hline
 \hline
\end{tabular}
\end{table}

\begin{table}[hb]
\centering
\caption{\label{tab:supp_photon_filtering} Entry in filtered dataset}
\begin{tabular}{|p{4cm}|p{12cm}|p{1.5cm}|}
\hline
Field name & Field description & Field data-type \\ \hline\hline
trigger index & Experiment Marker index preceding event in midpoint & uint64\\ \hline
node1 trigger timestamp & Absolute timestamp of corresponding Experiment marker on node1 & uint64\\ \hline
node2 trigger timestamp & Absolute timestamp of corresponding Experiment marker on node2 & uint64\\ \hline
detection bin index & Index of detection bin this event is detected in & uint32 \\ \hline
detA counts & Number of counts measured by nanowire A for this event & uint16\\ \hline
detB counts & Number of counts measured by nanowire B for this event & uint16\\ \hline
detA relative timestamp & Timestamp w.r.t. the start of detection bin of first count measured in nanowire A & int32\\ \hline
detB relative timestamp & Timestamp w.r.t. the start of detection bin of first count measured in nanowire B & int32\\ \hline 

 \hline
\end{tabular}
\end{table}

\begin{table}[hb]
\centering
\caption{\label{tab:supp_equipment_freqlock} Frequency lock hardware }
\begin{tabular}{|p{5cm}|p{5cm}|}
\hline
Equipment name & Part name/number \\ \hline\hline
Balanced photodetector & Thorlabs PDB482C-AC \\ \hline
Mixer & MiniCircuits ZX05-1l+    \\ \hline
Amplifier & Femto DHPVA-201 \\ \hline
Bandpass filter & MiniCircuits SBP-100+ \\ \hline
Downmix signal generator \SI{325}{\mega\hertz} & AnaPico APSin6010 \\ \hline
Beat reference generator \SI{100}{\mega\hertz} & AnaPico APSin6010 \\ \hline
Phase-Frequency-Detector & AnalogDevices HMC3716LP4E\\ \hline
Differential probe & Pintek DP-60HS\\ \hline
Track \& Hold & Texas Instruments OPA1S2384\\ \hline
Integrator &  NewFocus LB1005 \\ \hline
ADC, DAC & Analog Discovery 2 \\ \hline
\hline
\end{tabular}
\end{table}

\begin{table}[hb]
\centering
\caption{\label{tab:supp_overview_numerical_values} Numerical values of inputs to calculation and simulations}
\begin{tabular}{|p{2cm}|p{2.5cm}|p{2.5cm}|p{2.5cm}|p{2.4cm}|p{1.5cm}|p{2cm}|}
\hline
Window length [$ns$] & $p_1$ [$10^{-5}$] & $p_2$ [$10^{-5}$] & $p_{DC_A}$ [$10^{-6}$] & $p_{DC_B}$ [$10^{-6}$] & $C_{M}$ & $C_{dist}$\\ \hline\hline
6& $2.1\pm0.4$& $1.4\pm0.4$& $0.84\pm0.1$& $0.84\pm0.1$& $13\pm4$& $92.37\pm0.32$\\ \hline
8& $2.7\pm0.4$& $1.8\pm0.4$& $1.12\pm0.15$& $1.12\pm0.13$& $19\pm4$& $154.7\pm0.4$\\ \hline
10& $3.2\pm0.4$& $2.1\pm0.4$& $1.40\pm0.19$& $1.40\pm0.17$& $30\pm5$& $221.7\pm0.5$\\ \hline
12& $3.7\pm0.4$& $2.4\pm0.4$& $1.67\pm0.22$& $1.68\pm0.20$& $37\pm6$& $297.9\pm0.6$\\ \hline
14& $4.1\pm0.4$& $2.7\pm0.4$& $1.95\pm0.26$& $1.96\pm0.23$& $44\pm7$& $367.6\pm0.6$\\ \hline
16& $4.4\pm0.4$& $2.9\pm0.4$& $2.23\pm0.30$& $2.24\pm0.27$& $53\pm7$& $436.9\pm0.7$\\ \hline
18& $4.7\pm0.4$& $3.1\pm0.4$& $2.51\pm0.34$& $2.52\pm0.30$& $74\pm9$& $496.5\pm0.7$\\ \hline
20& $5.0\pm0.4$& $3.2\pm0.4$& $2.8\pm0.4$& $2.79\pm0.33$& $93\pm10$& $558.9\pm0.8$\\ \hline
22& $5.2\pm0.4$& $3.4\pm0.4$& $3.1\pm0.4$& $3.1\pm0.4$& $103\pm10$& $611.1\pm0.8$\\ \hline
24& $5.4\pm0.4$& $3.5\pm0.4$& $3.3\pm0.4$& $3.4\pm0.4$& $118\pm11$& $662.9\pm0.9$\\ \hline
26& $5.5\pm0.4$& $3.6\pm0.4$& $3.6\pm0.5$& $3.6\pm0.4$& $133\pm12$& $708.5\pm0.9$\\ \hline
28& $5.7\pm0.4$& $3.7\pm0.4$& $3.9\pm0.5$& $3.9\pm0.5$& $148\pm12$& $749.5\pm0.9$\\ \hline
30& $5.8\pm0.4$& $3.7\pm0.4$& $4.2\pm0.6$& $4.2\pm0.5$& $159\pm13$& $785.7\pm0.9$\\ \hline
\hline
\end{tabular}
\end{table}

\end{document}